# Advanced McMillan's equation and its application for the analysis of highly-compressed superconductors

E. F. Talantsev[1,2*]

[1] M.N. Mikheev Institute of Metal Physics, Ural Branch, Russian Academy of Sciences, 18, S. Kovalevskoy St., Ekaterinburg, 620108, Russia

[2] NANOTECH Centre, Ural Federal University, 19 Mira St., Ekaterinburg, 620002, Russia

[*] E-mail: evgeny.talantsev@imp.uran.ru

*Abstract*

A theory of electron-phonon mediated superconductivity requires the knowledge of full phonon spectrum, $\alpha^2(\omega) \cdot F(\omega)$, to calculate superconducting transition temperature, $T_c$. However, there is no experimental technique which can measure $\alpha^2(\omega) \cdot F(\omega)$ in highly-compressed near-room-temperature (NRT) superconductors to date. In this paper we propose to advance McMillan's approach (1968 *Phys Rev* **167** 331) which utilises the Debye temperature, $T_\theta$ (an integrated parameter of full phonon spectrum), that we deduced by the fit of experimentally measured temperature-dependent resistance data, $R(T)$, to Bloch-Grüneisen equation for highly-compressed black phosphorous, boron, GeAs, $SiH_4$, $H_xS$, $D_xS$, $LaH_x$ and $LaD_y$. By utilizing relations between $T_c$, $T_\theta$ and electron-phonon coupling strength constant, $\lambda_{e-ph}$ (which can be computed by first-principles calculations), it is possible to affirm/disprove the electron-phonon coupling mechanism in given superconductors. We show that computed $\lambda_{e-ph}$ for highly-compressed black phosphorous, boron, GeAs, $SiH_4$ and for one sample of $LaH_{10}$ are in a good agreement with $\lambda_{e-ph}$ values deduced from experimental data. It is also found remarkable constancy of $T_\theta = 1531 \pm 70\ K$ for $H_3S$ at different ageing stages. We also show that if phonon spectra of two isotopic counterparts have identical shape (or, in case of highly-compressed superconductor, the same material at different pressures), then within electron-phonon phenomenology these materials should obey the relation of $T_{\theta,1}/T_{\theta,2}=T_{c,1}/T_{c,2}=\omega_{ln,2}/\omega_{ln,2}$ (where subscripts 1 and 2 designate two isotopic counterparts). We report that for H$_3$S-D$_3$S pair ratios of $T_{c,H3S}/T_{c,D3S}=\omega_{ln,H3S}/\omega_{ln,D3S}=1.27$ are largely different from deduced $T_{\theta,H3S}/T_{\theta,D3S}=1.65$. This alludes that NRT superconductivity in H$_3$S-D$_3$S system is originated from more than one mechanism, where the electron-phonon coupling lifts $T_c$ in H$_3$S vs D$_3$S, but primary origin for NRT background of $T_c \sim 150$ K in both H$_3$S and D$_3$S remains to be discovered.

**1. Introduction.**

Demanded for several decades [1-4] near-room-temperature (NRT) superconductivity has been discovered by pioneer experimental work by Drozdov *et al* [5] who reported the superconducting transition temperature of $T_c = 203$ K in highly-compressed sulphur hydride, H$_3$S. To date, there is widely accepted point of view that the mechanism of NRT superconductivity in highly-compressed superhydrides/superdeuterides is the electron-phonon coupling [6-9]. If so, the superconducting transition temperature, $T_c$, and phonon spectrum characteristics (for instance, Debye frequency and Debye temperature, $\omega_D$ and $T_\theta$ respectively) should be linked through the electron-phonon coupling strength constant, $\lambda_{e-ph}$, introduced in Eliashberg theory [10]:

$$\lambda_{e-ph} = 2 \cdot \int_0^\infty \frac{\alpha^2(\omega) \cdot F(\omega)}{\omega} \cdot d\omega \qquad (1)$$

where ω is the phonon frequency, $F(\omega)$ is the phonon density of states, and $\alpha^2(\omega) \cdot F(\omega)$ is the electron-phonon spectral function (more details can be found elsewhere [11,12]). In this phenomenology, all superconductors are characterized as having weak ($\lambda_{e-ph} \ll 1$), intermediate ($\lambda_{e-ph} \sim 1$), and strong ($\lambda_{e-ph} \gg 1$) coupling.

For weak-coupled superconductors, Bardeen, Cooper and Schrieffer [13] proposed an expression which links $T_c$, $T_\theta$ and $\lambda_{e-ph}$:

$$T_c = T_\theta \cdot e^{-\left(\frac{1}{\lambda_{e-ph} - \mu^*}\right)} \quad (2)$$

where μ* is the Coulomb pseudopotential parameter, which is for a wide range of superconductors (including highly-compressed hydrides/deuterides) is assumed to be within a range of μ* = 0.10-0.17 [6].

Allen and Dynes [14,15] derived an asymptote for large $\lambda_{e-ph}$ values within the BCS theory [13]:

$$T_c = \frac{\lambda_{e-ph}}{2} \cdot T_\theta. \quad (3)$$

From Eq. 3 it is clear that $T_c$ (within the BCS theory [13]) can be "arbitrarily large" [15], because high $T_\theta$ is not necessarily a requirement for high $T_c$.

McMillan [16] performed advanced analysis of the problem by utilizing the Eliashberg theory [10] and proposed the equation:

$$T_c = \left(\frac{1}{1.45}\right) \cdot T_\theta \cdot e^{-\left(\frac{1.04 \cdot (1+\lambda_{e-ph})}{\lambda_{e-ph} - \mu^* \cdot (1+0.62 \cdot \lambda_{e-ph})}\right)} \quad (4)$$

which is highly accurate for a wide range of the coupling strength, $\mu^* \leq \lambda_{e-ph} \leq 1.5$ [14,15], and it is widely used to evaluate the $T_c$ in phonon mediated superconductors, including 2D superconductors [17].

It should be noted, that McMillan [16] introduce the multiplicative pre-factor of $\left(\frac{1}{1.45}\right) \cdot T_\theta$ in Eq. 4 based on experimental convenience to use the Debye temperature, $T_\theta$ (or, the Debye frequency, $\omega_D$), in comparison with more complicated functions which can be defined based on full $\alpha^2(\omega) \cdot F(\omega)$ phonon spectrum. And the reason for this was, that even for intrinsic uncompressed superconductors there are a very limited number of experimental techniques (for instance, the point-contact tunnelling spectroscopy) which can be used to measure full $\alpha^2(\omega) \cdot F(\omega)$ phonon spectrum. It should be stressed that none of these techniques have been ever applied for highly-compressed superconductors because of experimental challenges.

However, the Debye temperature, $T_\theta$, can be in "the safest" manner [18] deduced from the fit of experimental temperature dependent resistivity data, $\rho(T)$, to Bloch-Grüneisen (BG) equation [19,20]:

$$\rho(T) = \rho_0 + A \cdot \left(\frac{T}{T_\theta}\right)^5 \cdot \int_0^{\frac{T_\theta}{T}} \frac{x^5}{(e^x-1)\cdot(1-e^{-x})} \cdot dx \tag{5}$$

where, the first term is the residual resistivity due to the scattering of conduction charge carriers (holes) on the static defects of the crystalline lattice, while the second term describes the hole-phonon scattering, and $A$ and $T_\theta$ are free-fitting parameters.

Eqs. 4,5 are instructive tool to derive $\lambda_{e\text{-ph}}$ in intrinsic [21,22] and highly-compressed [23] superconductors, which works from weak to intermediate coupling strength of $\lambda_{e-ph} \lesssim 1$. In this paper, we further show that Eqs. 4,5 can be used to derive $\lambda_{e\text{-ph}}$ values in a variety of highly-compressed superconductors (i.e. black phosphorous [24,25], boron [26], GeAs [27], and silane [28]), which are in a good agreement with $\lambda_{e\text{-ph}}$ values reported by the first-principles calculations studies.

It should be noted, that one of the most widely used approximation of the Eliashberg theory [10] beyond Eq. 4 has been proposed by Allen and Dynes [14,15,18]:

$$T_c = \left(\frac{1}{1.20}\right) \cdot \left(\frac{\hbar}{k_B}\right) \cdot \omega_{ln} \cdot e^{-\left(\frac{1.04\cdot(1+\lambda_{e-ph})}{\lambda_{e-ph}-\mu^*\cdot(1+0.62\cdot\lambda_{e-ph})}\right)} \cdot f_1 \cdot f_2 \tag{6}$$

where $k_B$ is the Boltzmann constant, and $\hbar$ is reduced Planck constant, and:

$$\omega_{ln} = exp\left[\frac{\int_0^\infty \frac{ln(\omega)}{\omega}\cdot F(\omega)\cdot d\omega}{\int_0^\infty \frac{1}{\omega}\cdot F(\omega)\cdot d\omega}\right] \tag{7}$$

$$f_1 = \left(1 + \left(\frac{\lambda_{e-ph}}{2.46\cdot(1+3.8\cdot\mu^*)}\right)^{3/2}\right)^{1/3} \tag{8}$$

$$f_2 = 1 + \frac{\left(\frac{\langle\omega^2\rangle^{1/2}}{\omega_{ln}}-1\right)\cdot\lambda_{e-ph}^2}{\lambda_{e-ph}^2+\left(1.82\cdot(1+6.3\cdot\mu^*)\cdot\left(\frac{\langle\omega^2\rangle^{1/2}}{\omega_{ln}}\right)\right)^2}, \qquad (9)$$

$$\langle\omega^2\rangle^{1/2} = \frac{2}{\lambda_{e-ph}}\cdot\int_0^\infty \omega\cdot\alpha^2\cdot F(\omega)\cdot d\omega. \qquad (10)$$

where $f_1$ and $f_2$ are so-called the strong-coupling correction function and the shape correction function, respectively [15]. It should be stressed, that Eq. 6 is also an approximation of general Eliashberg theory [10], and Eq. 6 is not accurate approximation for some superconductors, which was discussed by Allen and Dynes [14,15,18].

Primary reason, why $\left(\frac{1}{1.45}\right)\cdot T_\theta$ (in Eq. 4) was proposed to replace by $\left(\frac{1}{1.20}\right)\cdot\omega_{ln}$ (in Eq. 6) is that the numerator $\left(\frac{1}{1.45}\right)$ in Eq. 4 was deduced for niobium (the element with highest $T_c$ and intermediate $\lambda_{e-ph}$). This numerator needs to be slightly adjusted from one material to another, while the pre-factor $\left(\frac{1}{1.20}\right)$ in Eq. 5 is much less varied (however, this does not mean that it should not need to be adjusted). Allen and Dynes [15] explored the dependence of $T_c$ vs $\omega_{ln}$ for a wide range of theoretically possible values of $0 \leq \lambda_{e-ph} \leq 10^6$, while the highest experimental $\lambda_{e-ph}$ value at the time of their work was $\lambda_{e-ph} = 2.59$ (this value is still one of the largest experimentally measured $\lambda_{e-ph}$ to date).

Based on a fact that modern first-principles calculations techniques can compute $\alpha^2(\omega)\cdot F(\omega)$ spectrum (and, thus, $\lambda_{e-ph}$) with a very high accuracy, Eq. 6 became a primary equation for the first-principles calculations studies for highly-compressed superhydrides/superdeuterides (extended reference list can be found elsewhere [6-9,29-32]). However, as we mentioned above, experimental measurements of $\alpha^2(\omega)\cdot F(\omega)$ in highly-compressed superconductors are not performed to date and alternative experimental approaches need to be developed for NRT superconductors in term to reveal the nature of the pairing mechanism in this compound.

In this paper, we propose one of these approaches which consider the isotopic counterparts (designated by subscripts 1 and 2) of one chemical compound. We propose the exact relation between $T_c$ and $T_\theta$ which is independent from particular $\alpha^2(\omega) \cdot F(\omega)$ spectrum. Truly, all first principles calculations studies show that the shape of $\alpha^2(\omega) \cdot F(\omega)$ spectra for isotopic counterparts of $H_3S$-$D_3S$ and of $LaH_{10}$-$LaD_{10}$ are practically undistinguishable, and thus, adjustable numerators of $\left(\frac{1}{1.45}\right)$ in Eq. 4 and of $\left(\frac{1}{1.20}\right)$ in Eq. 6 will be cancelled in the ratios:

$$\left.\frac{T_{c,1}}{T_{c,2}}\right|_{exp} = \left.\frac{\omega_{ln,1}}{\omega_{ln,2}}\right|_{first-principles\ calcs} = \left.\frac{T_{\theta,1}}{T_{\theta,2}}\right|_{exp\ \rho(T)}, \qquad (11)$$

where the subscript of *exp* $\rho(T)$ on the right part designates that $T_{\theta,1}$ and $T_{\theta,2}$ are deduced by the fit of experimental $\rho(T)$ data to Bloch-Grüneisen (BG) equation (Eq. 5).

It should be noted that Eq. 11 can be applied for any two superconductors which have the same or similar $\lambda_{e\text{-}ph}$ and $\mu^*$. In this regard, Eq. 11 is a new research tool to test the validity of electron-phonon mediated mechanism of superconductivity, which can be applied not only for two isotopic counterparts of the same chemical compound, but for any pair of superconductors for which computed, by the first principles calculations, $\lambda_{e\text{-}ph}$ and $\mu^*$ values will be close enough. Thus, surprisingly enough, the research task to reaffirm/disprove the electron-phonon coupling mechanism in highly compressed hydrides/deuterides is to compare experimentally observed $\left.\frac{T_{c,1}}{T_{c,2}}\right|_{exp}$ and/or computed $\left.\frac{\omega_{ln,1}}{\omega_{ln,2}}\right|_{first-principles\ calcs}$ (where the latter is, in fact, always calculated to match observed $T_{c,1}$ and $T_{c,2}$) with deduced $\left.\frac{T_{\theta,1}}{T_{\theta,2}}\right|_{exp\ \rho(T)}$.

By applying Eq. 11 to $H_3S$-$D_3S$ system, we find that the ratio of $\left.\frac{T_{\theta,H3S}}{T_{\theta,D3S}}\right|_{exp\ \rho(T)} \cong 1.65$ is remarkably different from $\left.\frac{T_{c,H3S}}{T_{c,D3S}}\right|_{exp} = \left.\frac{\omega_{ln,H3S}}{\omega_{ln,D3S}}\right|_{first-principles\ calcs} \cong 1.28$.

From this we conclude that there are two distinctive mechanisms which cause the raise of NRT superconductivity in highly-compressed $H_3S$-$D_3S$ system. One of the mechanism is the electron-phonon coupling which busts the observed $T_c$ in $H_3S$ vs $D_3S$, while the primary mechanism which is the origin for high background $T_c \gtrsim 150\ K$ in both isotopic counterparts remains to be discovered. This means that, due to observed $T_c$ in $D_3S$ is either not busted strongly, neither busted at all by the electron-phonon coupling, the research of $D_3S$ makes more fundamental interest in comparison with $H_3S$.

In this regard, the sulphur tritiate, $T_3S$, looks even more attractive to be studied, in comparison with $H_3S$ and $D_3S$, in term to reveal fundamental mechanism of NRT superconductivity, because of, at least, three reasons:

1. The contribution of electron-phonon coupling on the superconducting state in $T_3S$ should be even more supressed in comparison with $D_3S$. Thus, unrevealed yet background mechanism which causes the raise NRT superconductivity in highly compressed super-hydrides/deuterides will be less distorted by the electron-phonon interaction.

2. Much greater chemical activity of tritium in comparison with hydrogen and deuterium, which originates from low-level $\beta$-radioactivity of the former, can be served as additional catalytic factor (in addition to primary laser heating) to form $T_3S$ phase from $T_2S$ gas in diamond-anvil cell.

3. Ground state upper critical field, $B_{c2}(0)$, in $T_3S$ might be lower than ones in $H_3S$ and $D_3S$. This makes it possible to measure temperature dependent $B_{c2}(T)$ for $T_3S$ in a wider temperature range in comparison with $H_3S$ and $D_3S$, and, thus, to perform deeper analysis of $B_{c2}(T)$.

Despite a fact that the cost of $T_3S$ synthesis and studies will be much higher in comparison with ones for $H_3S$ and $D_3S$, the former looks to be a key material to reveal the mechanism of NRT superconductivity. Same approach is equally applied for other NRT superconductor systems including LaH$_x$-LaD$_y$-LaT$_z$ (however, for simplicity, we mention only H$_3$S-D$_3$S-T$_3$S system herein).

## 2. Utilized models

In our previous works [33-35], we analysed temperature dependent self-field critical current densities, $J_c$(sf,$T$), and the upper critical field, $B_{c2}(T)$, data in highly-compressed H$_3$S and showed that experiment supports the weak-coupling scenario in this superhydride. Kaplan and Imry [36] also showed that electron-phonon weak-coupling scenario is a valid theoretical model for compressed in H$_3$S. Thus, there is no a priory reason to reject the weak-coupling scenario from the consideration, and λ$_{e-ph}$ values derived by Eq. 2 will be designated as $\lambda_{e-ph,BSC}$. Deduced λ$_{e-ph}$ for strong-coupling scenario within BCS theory (Eq. 3) will be designated as $\lambda_{e-ph,BSC\ asymp}$, and coupling strength, $\lambda_{e-ph}$, computed by the first-principles calculations will be notated as $\lambda_{e-ph,fpc}$.

To turn now to strong-coupling scenario within the Eliashberg theory [10], we first should note that multiplicative shape correction function, $f_2$ (Eq. 8), has the term of $\left(\frac{\langle\omega^2\rangle^{1/2}}{\omega_{ln}}\right)$ which was never been experimentally measured in highly-compressed hydrides/deuterides to date. From other hand, all materials considered by Allen and Dynes (Table 1 [15]) have λ$_{e-ph}$ values within a range of $0.69 \leq \lambda_{e-ph} \leq 2.59$, which covers the range of λ$_{e-ph}$ reported for highly-compressed superhydrides/superdeuterides by first principles calculations studies [6-9,29-32]. For all these materials multiplicative shape correction function, $f_2$, can be, with accuracy of several percent, approximated by a parabolic function. We demonstrate this in Fig. 1, where $f_2$ function (Eq. 8) is calculated for μ* = 0.10, 0.15, 0.20 for all materials considered by Allen and Dynes (Table 1 [15]). The upper $\mu^* = 0.20$ bound is intentionally taken larger in Fig. 1 than conventionally used upper bound of $\mu^* = 0.15$.

Thus, with an accuracy of several percent, multiplicative shape correction function $f_2$ (Eq. 7) can be replaced by the parabolic function:

$$f_2^* = 1 + (0.0241 - 0.0735 \cdot \mu^*) \cdot \lambda_{e-ph}^2. \tag{12}$$

Numerators in Eq. 12 were deduced by the fit of tabulated $f_2$ values for all materials reported by Allen and Dynes (Table 1 [15]) for $\mu^* = 0.10, 0.12, 0.14, 0.16, 0.18, 0.20$.

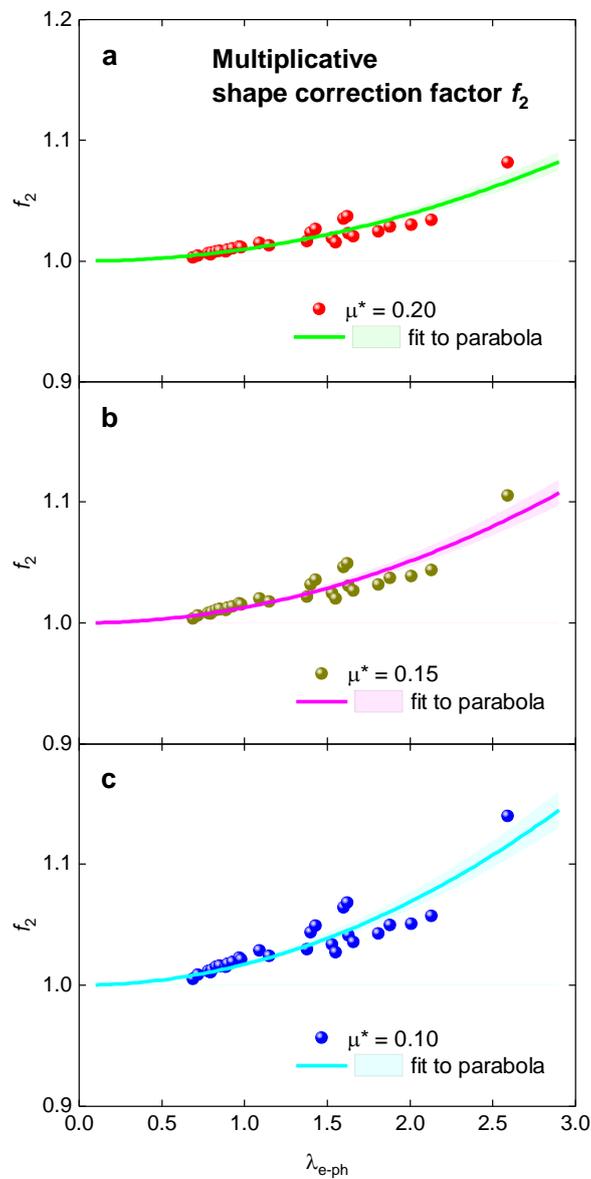

**Figure 1.** Multiplicative shape correction factor, $f_2$ (Eq. 9), for all materials considered by Allen and Dynes [10]. $f_2$ is calculated for $\mu^*=0.10$ (a), $\mu^*=0.15$ (b), and $\mu^*=0.20$ (c).

As this was mentioned by Allen [18], the "safest procedure" to extract $T_\theta$ from experimental data is to fit $\rho(T)$ to Eq. 10 which we employ in this paper. The coupling constant deduced by the advanced McMillan Eq. 4:

$$T_c = \left(\frac{1}{1.45}\right) \cdot T_\theta \cdot e^{-\left(\frac{1.04 \cdot (1+\lambda_{e-ph,aMcM})}{\lambda_{e-ph} - \mu^* \cdot (1+0.62 \cdot \lambda_{e-ph,aMcM})}\right)} \cdot f_1 \cdot f_2^* \tag{13}$$

will be designated as $\lambda_{e-ph,aMcM}$ as noted.

Considering a fact that all first-principles calculations studies show that the shape of phonon spectra, $\alpha^2(\omega) \cdot F(\omega)$ and $\lambda_{e-ph,fpc}$ for H$_3$S and D$_3$S [37] are practically undistinguishable from each other, one can write (the same consideration can be applied for other superconducting system, including LaH$_x$-LaD$_y$, but for simplicity we consider only H$_3$S-D$_3$S system):

$$\left.\frac{T_{c,H_3S}}{T_{c,D_3S}}\right|_{Eq.6} = \frac{\omega_{ln,H_3S} \cdot e^{-\left(\frac{1.04 \cdot (1+\lambda_{e-ph,H_3S})}{\lambda_{e-ph,H_3S} - \mu^* \cdot (1+0.62 \cdot \lambda_{e-ph,H_3S})}\right)} \cdot f_{1,H_3S} \cdot f_{2,H_3S}}{\omega_{ln,D_3S} \cdot e^{-\left(\frac{1.04 \cdot (1+\lambda_{e-ph,D_3S})}{\lambda_{e-ph,D_3S} - \mu^* \cdot (1+0.62 \cdot \lambda_{e-ph,D_3S})}\right)} \cdot f_{1,D_3S} \cdot f_{2,D_3S}} = \frac{\omega_{ln,H_3S} \cdot f_{1,H_3S} \cdot f_{2,H_3S}}{\omega_{ln,D_3S} \cdot f_{1,D_3S} \cdot f_{2,D_3S}} = \frac{\omega_{ln,H_3S}}{\omega_{ln,D_3S}}$$

(14)

$$\left.\frac{T_{c,H_3S}}{T_{c,D_3S}}\right|_{Eq.13} = \frac{T_{\theta,H_3S} \cdot e^{-\left(\frac{1.04 \cdot (1+\lambda_{e-ph,aMcM,H_3S})}{\lambda_{e-ph,1} - \mu^* \cdot (1+0.62 \cdot \lambda_{e-ph,aMcM,H_3S})}\right)} \cdot f_{1,H_3S} \cdot f_{2,H_3S}^*}{T_{\theta,D_3S} \cdot e^{-\left(\frac{1.04 \cdot (1+\lambda_{e-ph,aMcM,D_3S})}{\lambda_{e-ph,2} - \mu^* \cdot (1+0.62 \cdot \lambda_{e-ph,aMcM,D_3S})}\right)} \cdot f_{1,D_3S} \cdot f_{2,D_3S}^*} = \frac{T_{\theta,H_3S} \cdot f_{1,H_3S} \cdot f_{2,H_3S}^*}{T_{\theta,D_3S} \cdot f_{1,D_3S} \cdot f_{2,D_3S}^*} = \frac{T_{\theta,H_3S}}{T_{\theta,D_3S}}$$

(15)

$$\left.\frac{T_{c,H_3S}}{T_{c,D_3S}}\right|_{exp} = \left.\frac{\omega_{ln,H_3S}}{\omega_{ln,D_3S}}\right|_{first-principles\ calcs} = \left.\frac{T_{\theta,H_3S}}{T_{\theta,D_3S}}\right|_{exp\ \rho(T)}, \tag{16}$$

because as we show in Fig. 3, $f_2$ function can be approximated by parabolic function $f_2^*$ (Eq. 12).

Thus, major problem of the McMillan [16] approach (which is exact value for multiplicative numerator in Eq. 5), and of the Allen-Dynes approach [14,15,18] (which is experimentally unknown value of ω$_{ln}$ (Eq. 7)) are cancelled out for H$_3$S and D$_3$S counterparts (Eq. 16), and if the superconductivity in highly-compressed H$_3$S-D$_3$S mediates by the electron-phonon interaction, then three ratios: experimentally observed $\frac{T_{C,H_3S}}{T_{C,D_3S}}$, computed $\frac{\omega_{ln,H_3S}}{\omega_{ln,D_3S}}$ and derived $\frac{T_{\theta,H_3S}}{T_{\theta,D_3S}}$ should be equal to each other. It should be noted, that if isotopic counterparts for some compounds have different $\alpha^2(\omega) \cdot F(\omega)$ shape, then Eqs. 11,16 can be only approximately satisfied. However, taking in account that $f_2$ and $f_2^*$ are very slow function of $\lambda_{e-ph}$, Eqs. 11,16 can be still considered as the first order approximation.

We should stress that Eqs. 11,16 are free from the uncertainty related to the accuracy of multiplicative numerator of $\left(\frac{1}{1.20}\right)$ in Eq. 6, proposed by Allen and Dynes as the first order approximation which compromised between the complexity and the accuracy for the model. Because more thorough consideration requires the use of high-order momentums of normalized weight function:

$$\langle \omega^n \rangle = \frac{\int_0^\infty \omega^{n-1} \cdot \alpha^2(\omega) \cdot F(\omega) \cdot d\omega}{\int_0^\infty \frac{\alpha^2(\omega) \cdot F(\omega)}{\omega} d\omega}, \tag{16}$$

and/or, its mean values:

$$\overline{\omega^n} = \langle \omega^n \rangle^{1/n} = \left(\frac{\int_0^\infty \omega^{n-1} \cdot \alpha^2(\omega) \cdot F(\omega) \cdot d\omega}{\int_0^\infty \frac{\alpha^2(\omega) \cdot F(\omega)}{\omega} d\omega}\right)^{1/n}, \tag{17}$$

which, however (from the author's knowledge) these higher-orders momentums have been never implemented for the analysis.

To confirm that our approach (i.e., Eqs. 5,13) is an instructive tool to analyse experimental data for highly-compressed superconductors, in Section 4.3 we perform the comparison of deduced parameters deduced by the *R*(*T*) analysis and by the upper critical field, $B_{c2}(T)$, analysis for highly-compressed GeAs (*P* = 15.3 GPa) recently reported by Liu *et al* [27].

To analyse $B_{c2}(T)$ we use a model [35]:

$$B_{c2}(T) = \frac{\phi_0}{2\cdot\pi\cdot\xi^2(0)} \cdot \left[\left(\frac{1.77-0.43\cdot\left(\frac{T}{T_c}\right)^2+0.07\cdot\left(\frac{T}{T_c}\right)^4}{1.77}\right)^2 \cdot \frac{1}{1-\frac{1}{2\cdot k_B\cdot T}\int_0^\infty \frac{d\varepsilon}{\cosh^2\left(\frac{\sqrt{\varepsilon^2+\Delta^2(T)}}{2\cdot k_B\cdot T}\right)}}\right] \quad (18)$$

where $\phi_0 = 2.07\cdot 10^{-15}$ Wb is flux quantum, $\xi(0)$ is the ground state coherence length, , and temperature dependent amplitude of the energy gap, $\Delta(T)$, is taken from Gross *et al* [38]:

$$\Delta(T) = \Delta(0) \cdot \tanh\left[\frac{\pi\cdot k_B \cdot T_c}{\Delta(0)} \cdot \sqrt{\eta \cdot \frac{\Delta C}{C} \cdot \left(\frac{T_c}{T}-1\right)}\right] \quad (19)$$

where $\Delta C/C$ is the relative jump in electronic specific heat at $T_c$, and $\eta = 2/3$ for *s*-wave superconductors [38]. In result, four fundamental parameters of a superconductor, i.e. $\xi(0)$, $\Delta(0)$, $\Delta C/C$ and $T_c$, can be deduced by fitting experimental $B_{c2}(T)$ data to Eq. 18. We need to clarify that $\xi(0)$ determines the ground state $B_{c2}(0)$ amplitude, while $\Delta(0)$ and $\Delta C/C$ are deduced from the shape of $B_{c2}(T)$ curve (which is the part of Eq. 18 in square brackets). BCS weak coupling limits are:

$$\frac{2\cdot\Delta(0)}{k_B\cdot T_c} = 3.53, \quad (20)$$

$$\frac{\Delta C}{C} = 1.43, \quad (21)$$

and, thus, deduced values of $\frac{2\cdot\Delta(0)}{k_B\cdot T_c}$ and $\frac{\Delta C}{C}$ are independent quantities which can characterize the coupling strength in addition to $\lambda_{e-ph,BCS}$ and $\lambda_{e-ph,aMcM}$ values.

We should stress that Eqs. 18,19 are based on the assumption that the amplitude and the phase coherence have established in the material, and thus superconducting condensate has forms. This means that Eqs. 18,19 are applied for the state when:

$$\rho(T) = 0\ \Omega\cdot m \quad (22)$$

In other words, Eq. 18 is applied when the London penetration depths, $\lambda(T)$, the superconducting coherence length, $\xi(T)$, and the Ginzburg-Landau parameter $\kappa(T) = \frac{\lambda(T)}{\xi(T)}$ have finite values:

$$\begin{cases} \lambda(T) \neq \infty \\ \xi(T) \neq \infty \\ \kappa(T) \text{ defined} \end{cases} \quad (23)$$

This means that experimental data which are valid to be fitted by Eq. 18 should be obey the $B_{c2}(T)$ definition based on Eq. 22 criterion. We note, that the upper critical field, $B_{c2}(T)$, is very often defined by 50% fraction of the normal state resistivity, $\rho_{norm}(T)$, criterion:

$$\rho(T) = 0.50 \cdot \rho_{norm}(T) \quad (24)$$

which, cannot be, rigorously speaking, be fitted by Eq. 18, except if the width of the superconducting transition, $\Delta T_c$ (which can be defined as the temperature range where 90% and 10% fractions of $\rho_{norm}(T)$ achieve), is very narrow, $\Delta T_c \ll T_c$. Truly, 50% drop in the resistivity at some temperature $T$ (we consider the case, when $\rho(T) = 0 \; \Omega \cdot m$ achieves at lower $T$) indicates that there are large superconducting order parameter fluctuations (in space and time), at which, however, neither $\lambda(T)$, nor $\xi(T)$ are defined yet, because the phase coherence for the condensate has not been formed at this temperature. This issue is closely related to the definition of the superconducting transition temperature, $T_c$, which we discuss in Section 3. We note, that $B_{c2}(T)$ defined by Eq. 22, is also referred as the irreversibility field, $B_{irr}(T)$, especially for cuprate superconductors.

Based on all above, the validity of electron-phonon mediated mechanism of superconductivity in highly-compressed superhydrides/superdeuterides can be, surprisingly enough, affirmed/disproved by the deducing the ratio of the Debye temperatures, $\frac{T_{\theta,H_3S}}{T_{\theta,D_3S}}$, for isotopic counterparts, and compare this ratio with the ratio of experimentally observed $\left.\frac{T_{c,H_3S}}{T_{c,D_3S}}\right|_{exp}$. We report on the result of these studies for $H_3S$-$D_3S$ system, herein.

## 3. $T_c$ definition

Due to primary focus of this paper is to compare experimentally observed and computed $T_c$ values, there is a need to make strict definition for the superconducting transition temperature, which will be in use herein. In some papers, $T_c$ is defined at temperature of the 95%, 50%, 10%, or other fractions of the normal state resistivity, $\rho_{norm}(T)$, while the most rigorous definition is at the zero resistivity point, $\rho(T) = 0\ \Omega\cdot m$ (Eq. 24).

The definition of $T_c$ as the fraction of the $\rho_{norm}(T)$ in highly-compressed hydrides/deuterides is widely used, however, this definition unavoidably raises many problems. For instance, a drop in $\rho(T) = 0.95\cdot\rho_{norm}(T)$ at some temperature $T$ cannot be a definitive confirmation for the superconducting state, before either $\rho(T) = 0\ \Omega\cdot m$, either Meissner-Ochsenfeld diamagnetic response can be detected in experiment. To demonstrate this problem, in Fig. 2 we show $\rho(T)/\rho(T = 270\ K)$ data for highly-compressed YH$_3$ ($P = 44$ GPa) reported by Matsuoka *et al* [39].

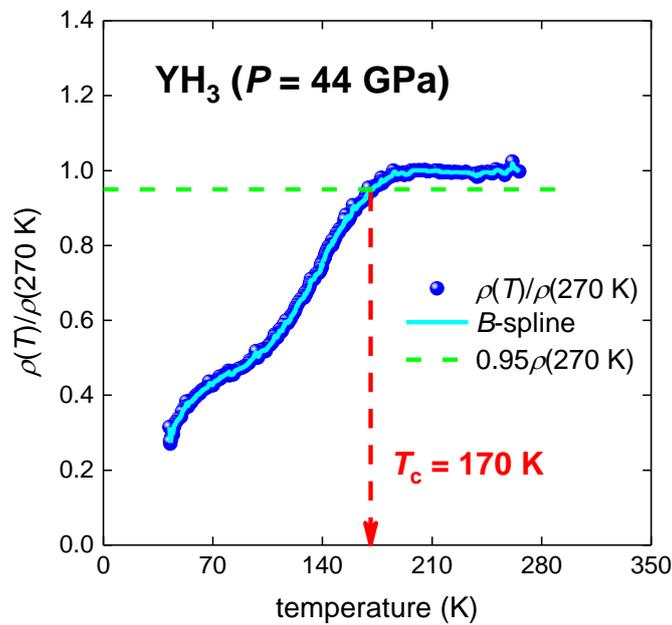

**Figure 2.** Temperature dependent reduced resistivity $\rho(T)/\rho(T = 270\ K)$ for highly compressed YH$_3$ (raw data reported by Matsuoka *et al* [39]) where $T_c$ is defined by $\rho(T)/\rho(T = 270\ K) = 0.95$ criterion.

If one applies the criterion of $\rho(T)/\rho(T = 270\text{ K}) = 0.95$, the superconducting transition temperature can be defined as high as $T_c = 170$ K. This means, that if $T_c$ definition is based on any fraction of $\rho_{norm}(T)$, i.e., $\rho_{norm}(T) \neq 0\ \Omega \cdot m$, the story of NRT superconductivity should be started in 2007, when Matsuoka *et al* [39] had reported their $\rho(T)$ dataset for highly compressed YH$_3$. We note, that recently NRT superconductivity with $T_c = 218$-$243$ K (defined by $\rho(T) = 0.99 \cdot \rho_{norm}(T)$ criterion) in yttrium hydrides compressed at $P = 165$-$237$ GPa has been reported by Troyan *et al* [31] and Kong *et al* [40]. Based on the criterion of $\rho(T) = 0.99 \cdot \rho_{norm}(T)$, the superconducting transition temperature in the yttrium hydride at $P = 44$ GPa [39], $T_c = 182$ K, is not much different from recently reported values of $T_c = 218$-$243$ K [31,40]. We, however, should stress that as Troyan *et al* [31], as Kong *et al* [40] reported experimental $\rho(T)$ curves which have reached zero resistivity point, $\rho(T) = 0\ \Omega \cdot m$, within a narrow superconducting transition width, $\Delta T_c$.

To further demonstrate that the definition of $T_c$ is crucially important, we can refer recent report by Cao *et al* [41] who discovered superconducting state in the magic-angle twisted bilayer graphene (MATBG) (i.e., 2D sheet where two single layers of graphene are rotated at Moiré superlattice angle, θ). In Fig. 3 we show raw $R(T)$ curve for MATBG sample with twisted angle of $\theta = 1.05°$ (raw data is from Ref. [41]). Three lines in the Fig. 3 indicate $R(T)/R_{norm}(T) = 1.0; 0.90; 0.50$ criteria and red data point indicate zero resistance state of $R(T = 0.9\ K) \cong 0\ \Omega$.

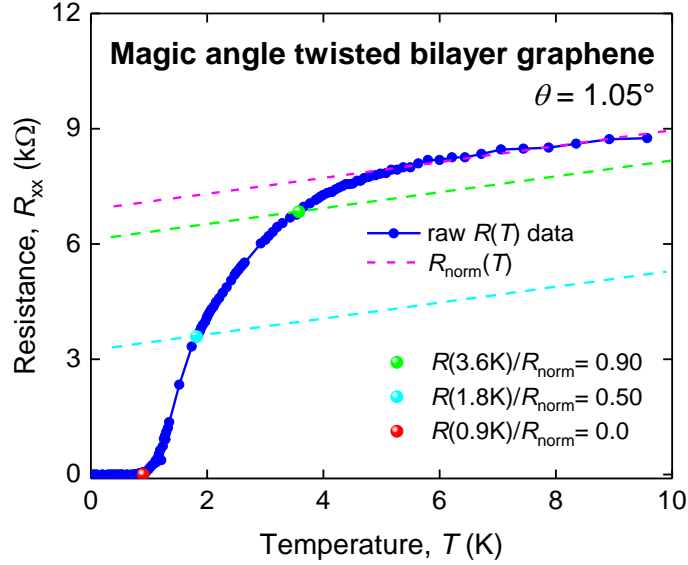

**Figure 3.** Temperature dependent resistivity $R(T)$, $R(T)/R_{norm}(T) = 1.0; 0.90; 0.50$ lines and $R(T = 0.9\ K)/R_{norm}(T) = 0.0$ data point are shown for MATBG ($\theta = 1.05°$) (raw data reported by Cao *et al* [41]).

Cao *et al* [41] used $T_c$ definition of $R(T)/R_{norm}(T) = 0.5$ and defined $T_c \cong 1.7\ K$. This definition places MATBG in near proximity to the Bose-Einstein condensate materials (i.e., $^4$He, $^6$Li, $^{40}$K) in the Uemura plot. And also, this was the primary reason to claim non-phonon mediated mechanism of superconductivity in MATBG. However, the analysis of the self-field critical current densities, $J_c(sf,T)$ [42], as well as *ab initio* calculations [17], showed that the superconductivity in MATBG is phonon mediated and $T_c$ should be defined by the temperature at which condensate phase coherence is established (i.e., by $R(T \cong 1.2\ K) \rightarrow 0\ \Omega$ criterion). This $T_c$ definition places MATBG (in the Uemura plot) in the band where all cuprates, pnictides, fullerens, heavy fermions and highly-compressed hydries/deuterides are located [43,44].

However, for highly-compressed superconductors, the definition of $\rho(T=T_c) = 0\ \Omega \cdot m$, unfortunately cannot be used either, because many $\rho(T)$ curves never reach $\rho(T) = 0\ \Omega \cdot m$, due to experimental challenges (for instance, we can refer $\rho(T)$ data for highly-compressed boron [26] and lithium [45]). In some experiments, if even $\rho(T) = 0\ \Omega \cdot m$ has been reached, the transition width are wide, $\Delta T_c \sim (0.3-0.5) \cdot T_c$.

Based on all above, it should be stressed that the comparison of computed (by the first-principles calculations) and experimentally measured $T_c$ values has, unfortunately, a very large uncertainty, which is based on the chosen criterion of $T_c$ definition. Due to $T_c$ is strongly linked to $\lambda_{e\text{-ph}}$, the robustness of computed $\lambda_{e\text{-ph}}$, $T_c$ and $\omega_{ln}$ values by the first-principles calculations is largely unknown. We demonstrate this issue by the analysis of experimental data for highly-compressed elemental boron (Sec. 4.2) and GeAs (Sec. 4.3).

In this paper, by compromising between all possible complications, we define $T_c$ by the criterion of $\rho(T) = 0$ $\Omega\cdot$m. If experimental $\rho(T)$ dataset does not reach zero resistivity point, $T_c$ will be defined either by the lowest experimentally available temperature, either (in a case, when $\rho(T)$ is flatten after the drop) by the temperature of the inflection point. For $T_c$ values defined by two latter criteria we report the ratio of $\rho(T)/\rho_{norm}(T)$, where $\rho_{norm}(T)$ is the extrapolative curve obtained by the $\rho(T)$ data fit to Bloch-Grüneisen equation (Eq. 5).

## 4. Results and Discussion

In this section we first show that the approach to use Eqs. 5,13 is a valid research tool to study highly-compressed superconductors. To demonstrate this, we perform $\rho(T)$ analysis and deduce $\lambda_{e-ph}$ values for highly-compressed black phosphorous ($P = 15$ GPa) [25] (Section 4.1), elemental boron ($P = 240$ GPa) [26] (Section 4.2), GeAs ($P = 15.3, 20.6$ and $24$ GPa) [27] (Sections 4.3), and silane ($P = 240$ GPa) [28] (Sections 4.4). Data for $H_3S$ and $D_3S$ are analysed in Section 4.5. The isotope effect in $H_3S$-$D_3S$ system is discussed in Section 4.6. For the analysis we draw largely on experimental data reported by M. I. Eremets group (Max-Planck Institut für Chemie, Mainz, Germany).

Raw experimental $R(T)$ data for $LaH_{10}$ and $LaD_{10}$ superconductors was kindly provided by Dr. M. I. Eremets and Dr. V. S. Minkov (Max-Planck Institut für Chemie, Mainz,

### 4.1. Black phosphorous compressed at $P$ = 15 GPa

Wittig and Matthias [24] discovered pressure-induced superconductivity in black phosphorous with $T_c$ = 4.7 K ($P \sim 10$ GPa). Recent experimental studies [46,47] of highly-compressed black phosphorous start to confirm conceptual idea of Hirsch [48,49] that high-temperature superconductivity should be considered as an effect of the interaction between positively charge carriers (holes) and vibrations of positively charged lattice ions (phonons), which is fundamentally different from the idea of BCS theory which considers the interaction between negatively charged carriers (electrons) and vibrations of positively charged lattice ions (phonons). Experiments showed [46,47] that there is clear correlation between $T_c$ and the charge carriers sign in highly-compressed black phosphorous.

In Figure 3 we show raw experimental $\rho(T)$ data for highly-compressed black phosphorous ($P$ = 15 GPa) reported by Shirotani *et al* [25] in their Figure 5. The $\rho(T)$ data fit to Eq. 5 is excellent and deduced $T_\theta$ and $\lambda_{e\text{-ph}}$ are presented in Table 1. To derive $\lambda_{e\text{-ph}}$ values we use conventional lower bound of $\mu^* = 0.10$ and the upper bound of $\mu^* = 0.17$ (for which the first-principles calculations results were reported by Li *et al* [47]).

**Table 1.** Deduced $T_\theta$ and calculated $\lambda_{e\text{-ph}}$ for highly-compressed black phosphorous at $P$ = 15 GPa.

| $T_c$ (K) | $T_\theta$ (K) | $\lambda_{e-ph,BCS\ asymp}$ | Assumed $\mu^*$ | $\lambda_{e-ph,BCS}$ | $\lambda_{e-ph,aMcM}$ | $\lambda_{e-ph,fpc}$ [47] |
|---|---|---|---|---|---|---|
| 6.7 | 563 ± 16 | 0.019 ± 0.001 | 0.10 | 0.314 ± 0.001 | 0.488 ± 0.003 | |
| | | | 0.17 | 0.384 ± 0.001 | 0.628 ± 0.005 | 0.627 - 0.673 |

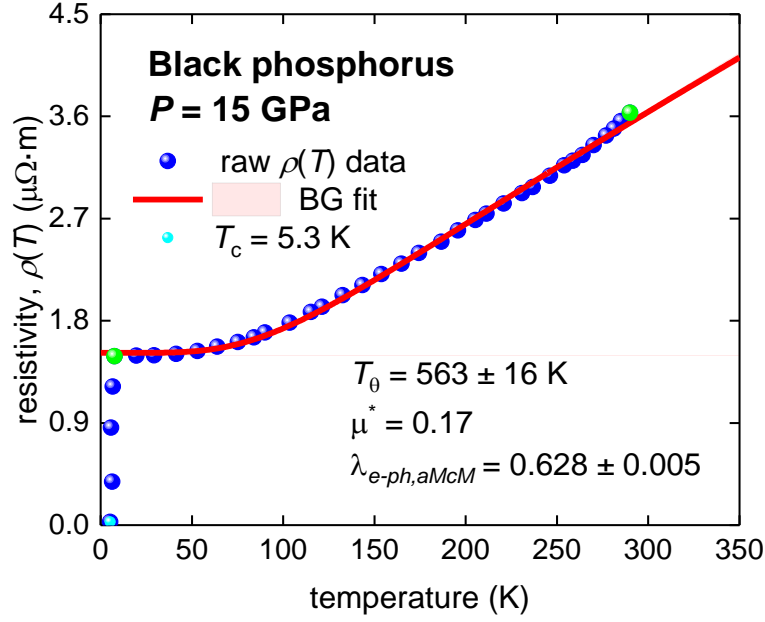

**Figure 4.** Resistivity data, $\rho(T)$, and fit to BG model (Eq. 5) for highly-compressed black phosphorus (raw data is from Ref. 47). The fit quality is $R = 0.998$. 95% confidence bar is shown. Green balls indicate the bounds for which $\rho(T)$ data was used for the fit to Eq. 5. Cyan ball shows $T_c$ defined by $\rho(T)/\rho_{norm}(T) = 0.80$ criterion.

It can be seen (Table 1) that deduced $\lambda_{e-ph,BCS} = 0.31 - 0.38$ is within BCS weak coupling limit, which originates from low value for the ratio of $\frac{T_c}{T_\theta} \cong 0.009$ (it can be noted that weak-coupling aluminium has $\lambda_{e-ph,BCS} = 0.30 - 0.38$ [18]).

Thus, there is an uncertainty, why more complicated intermediate coupling strength scenario should be considered, if weak-coupling limit of BCS is well satisfied. However, due to Li *et al* [47] performed first-principles calculations and reported that highly-compressed black phosphorus has $\lambda_{e-ph,fpc} = 0.63 - 0.67$ (µ* = 0.17) we calculated $\lambda_{e-ph,aMcM} = 0.63$ (µ* = 0.17) by our Eq. 13, which appears to be in excellent agreement with the value reported by Li *et al* [47].

This result shows that calculated coupling strength $\lambda_{e-ph}$ is actually dependent on the chosen model, rather than to be unique characteristic value for given superconductor. The only valid result, we can trust so far, that because of calculated $\lambda_{e-ph,BCS\ assymp} = 0.019 \ll$ $\mu^*$, then phonon-mediated Cooper pairs cannot exist in the assumption of strong-coupling scenario (which reflects a simple fact that there are no reasons to assume that $\lambda_{e-ph}$ can be large, if the superconductor has $\frac{T_c}{T_\theta} \cong 0.009$).

### 4.2. Elemental boron compressed at $P$ = 240 GPa

Eremets *et al* [26] discovered that elemental boron transforms into superconductor with $T_c > 4\ K$ at pressure of $160\ GPa \leq P \leq 250\ GPa$. Ma *et al* [50] calculated $\lambda_{e-ph,fpc} = 0.38 - 0.39$ for $\mu^* = 0.12$ in the wide pressure range of $160\ GPa \leq P \leq 273\ GPa$. Several years later, Sun *et al* [51] studied highly-compressed boron nanowires and reported $T_c = 1.5\ K$ at pressure of $P$ = 84 GPa, and Shimizu *et al* [52] confirmed the superconducting state in bulk boron with $T_c \sim 3\ K$ at pressure of $P = 159 \pm 5\ GPa$.

Eremets *et al* [26] in their Figure 2 reported experimental resistance data, $R(T)$, for the boron at different pressures, from which in Figure 4 we show $R(T)$ at $P$ = 240 GPa. It can be seen that $R(T) = 0\ \Omega$ has not been achieved and thus we use a series of $T_c$ definitions, $R(T)/R_{norm}(T) = 0.95; 0.80; 0.67$, where the latter is defined by the inflection point.

The $R(T)$ data fit to Eq. 5 is excellent and deduced $T_\theta = 314 \pm 2\ K$. In Table 2 we show deduced $\lambda_{e-ph}$ values in the assumption of $\mu^* = 0.12$ (which is chosen to be the same with one used by Ma *et al* [50]).

**Table 2.** Deduced $T_\theta$ and calculated $\lambda_{e-ph}$ for highly-compressed boron at $P$ = 240 GPa with assumed $\mu^* = 0.12$.

| $T_c$ (K) | $T_\theta$ (K) | $\lambda_{e-ph,asymp}$ | $\lambda_{e-ph,BCS}$ | $\lambda_{e-ph,aMcM}$ | $\lambda_{e-ph,fpc}$ [50] |
|---|---|---|---|---|---|
| 10.1<br>$R(T)/R_{norm}(T) = 0.95$ |  | 0.064 | 0.411 ± 0.001 | 0.769 ± 0.002 |  |
| 8.9<br>$R(T)/R_{norm}(T) = 0.80$ | 314 ± 2 | 0.057 | 0.401 ± 0.001 | 0.731 ± 0.002 | 0.39<br>($P$ = 215-279 GPa) |
| 7.4<br>$R(T)/R_{norm}(T) = 0.67$ |  | 0.050 | 0.387± 0.001 | 0.683 ± 0.002 |  |

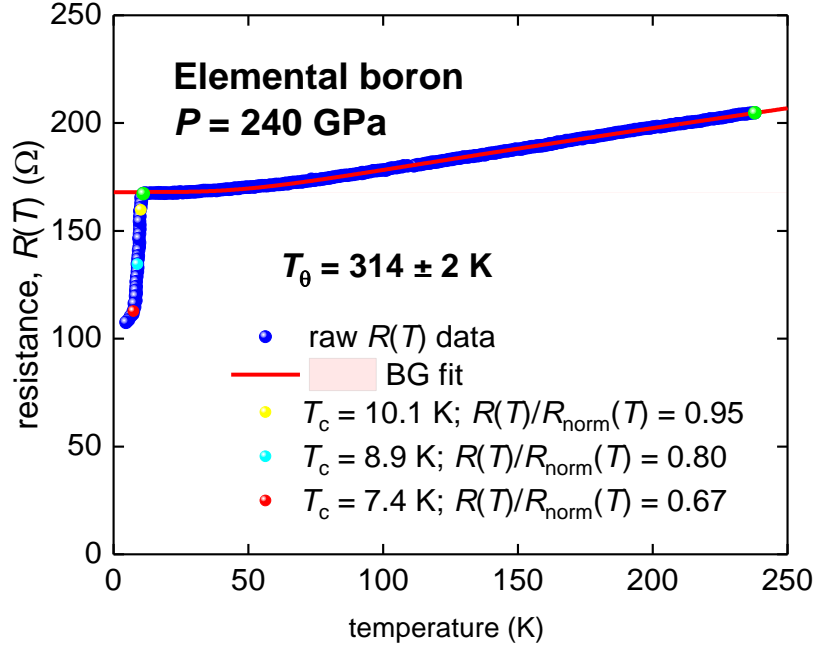

**Figure 5.** Resistance data, $R(T)$, and fit to BG model (Eq. 5) for highly-compressed elemental boron (raw data is from Ref. 26). The fit quality is $R = 0.9994$. 95% confidence bar is shown. Green balls indicate the bounds for which $\rho(T)$ data was used for the fit. Yellow ball shows $T_c$ defined by $R(T)/R_{norm}(T) = 0.95$ criterion; cyan ball shows $T_c$ defined by $R(T)/R_{norm}(T) = 0.80$ criterion; and red ball shows $T_c$ defined by $R(T)/R_{norm}(T) = 0.67$ criterion.

It should be stressed (Table 2), that $T_c$ values defined by three different criteria are varied by 36%, however this a large variation in $T_c$ corresponds to very small changes in $\lambda_{e-ph,BCS}$ (6%) and $\lambda_{e-ph,aMcM}$ (13%). Based on computed value of $\lambda_{e-ph,fpc} = 0.39$ [50] one can conclude that highly-compressed elemental boron ($P = 240$ GPa) is a weak-coupling superconductor. This conclusion is also supported by low value for the ratio of $\frac{T_{c,0.67R_{norm}}}{T_\theta} = \frac{7.4\ K}{314\ K} \lesssim 0.024$, and $\frac{T_{c,0.95R_{norm}}}{T_\theta} = \frac{7.4\ K}{314\ K} \lesssim 0.032$.

However, there is a question, how the accuracy of first-principles calculations and numerators in Eqs. 4,6 can be evaluated, if the change in $T_c$ by more than $\frac{1}{3}$ corresponds to a minor changes in computed $\lambda_{e-ph}$. This means that numerators in Eqs. 4,6 (i.e. $\frac{1}{1.45}$ and $\frac{1}{1.20}$, correspondingly) practically do not make any effect on calculated $T_c$ values.

## 4.3. Highly-compressed superconducting GeAs

Liu *et al* [27] reported that superconducting state can be induced in semiconducting compound of GeAs at pressures $P \geq 10\ GPa$. For GeAs subjected to pressure of $P = 15.3\ GPa$, Liu *et al* [27] reported $R(T,B)$ curves (Fig. 4(a) [27]) from which $B_{c2}(T)$ was deduced by the criterion of:

$$R(T, B_{c2}(T)) = 0.01 \cdot R_{norm}(T = 9\ K) \tag{25}$$

Fit of $B_{c2}(T)$ to Eq. 18 is shown in Fig. 6. Deduced parameters (within uncertainties) are very close to weak-coupling limits of BCS theory:

$$\frac{2 \cdot \Delta(0)}{k_B \cdot T_c} = 3.55 \pm 0.23 \tag{26}$$

$$\frac{\Delta C}{C} = 1.56 \pm 0.32 \tag{27}$$

Raw $R(T)$ curves (from Fig. 3(a) of Ref. 27) and data fits to BG equation for GeAs compressed at $P$ = 15.3, 20.6, and 24.0 GPa are shown in Fig. 7, where $R(T)/R_{norm}(T) = 0.0$ points are taken from Fig. 3(b) [27], and $R(T)/R_{norm}(T) = 0.95$ for Fig. 3(a) [27]. Deduced $\lambda_{e-ph}$ values in assumption of $\mu^* = 0.10$ (which is the same as one used by Liu *et al* [27]) are shown in Table 3.

It should be noted (Table 2), that at $P$ = 24 GPa, $T_c$ values defined by $R(T)/R_{norm}(T) = 0.0$ and $R(T)/R_{norm}(T) = 0.95$ criteria are varied by nearly twice (i.e. 96%), however this large variation in $T_c$ corresponds to a small changes in $\lambda_{e-ph,BCS}$ (10%) and $\lambda_{e-ph,aMcM}$ (17%).

It can be seen also (Table 3), that calculated $\lambda_{e-ph,fpc} = 0.51$ (at $P$ = 24 GPa) neither corresponds to the $T_c$ = 2.5 K (defined by $R(T)/R_{norm}(T) = 0.0$ criterion), nor by $T_c$ = 4.9 K (defined by $R(T)/R_{norm}(T) = 0.95$). Computed $\lambda_{e-ph,fpc} = 0.51$ corresponds to $T_c$ = 4.46 K which means that the criterion for Tc defined by $R(T)/R_{norm}(T) = 0.81$. However, there are no sensible explanations why $R(T)/R_{norm}(T) = 0.81$ criterion should be valid.

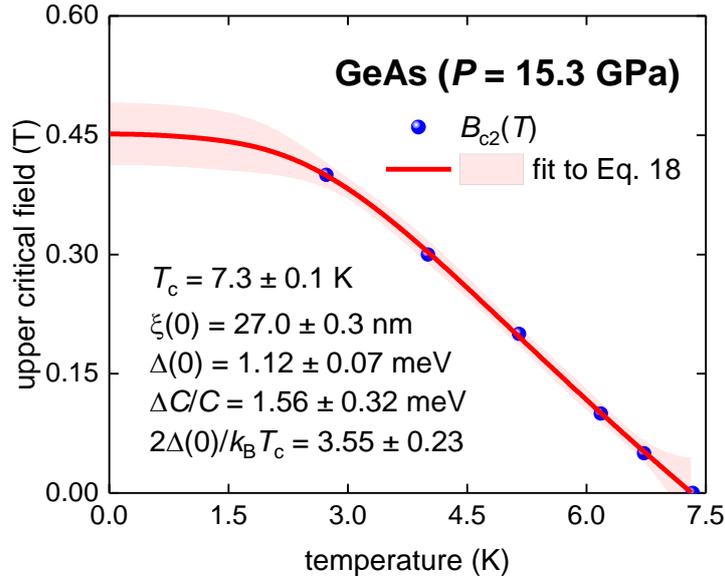

**Figure 6.** The fit to Eq. 18, the upper critical field, $B_{c2}(T)$, for GeAs compressed at pressure of $P = 15.3$ GPa (raw data is from Ref. 27), the fit quality is $R = 0.9998$, 95% confidence bar is shown.

Similarly, at $P = 20.6$ GPa, computed $\lambda_{e-ph,fpc} = 0.51$ corresponds to $T_c = 4.54$ K which corresponds to $R(T)/R_{norm}(T) = 0.83$ criterion, which is also (despite its proximity to $R(T)/R_{norm}(T) = 0.81$ criterion deduced for GeAs compressed at $P = 24$ GPa) cannot be justified by any sensible background physics (at least, for the current status of the theory of superconductivity).

However, one can see in Table 3, that deduced $\lambda_{e-ph,BCS} \cong 0.30 - 0.36$ (which are within BCS weak-coupling limit) are well matched $\frac{2 \cdot \Delta(0)}{k_B \cdot T_c} = 3.55 \pm 0.23$ and $\frac{\Delta C}{C} = 1.56 \pm 0.32$ values (which are also within BCS weak-coupling limits) deduced from the analysis of $B_{c2}(T)$ data.

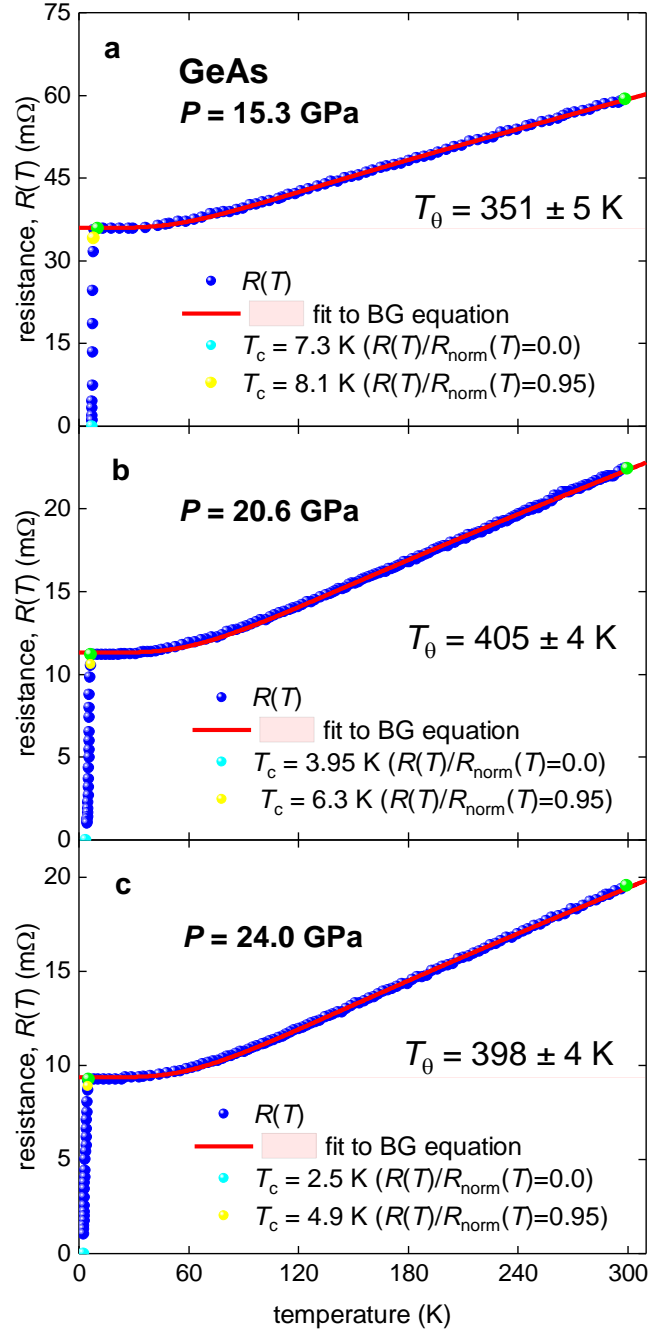

**Figure 7.** $R(T)$ data and fits to BG model (Eq. 5) for highly-compressed GeAs (raw $R(T)$ data is from Ref. 27). 95% confidence bars are shown. Green balls indicate the bounds for which $\rho(T)$ data was used for the fit. (a) $P = 15.3$ GPa ($T_\theta = 351 \pm 5$ K), fit quality R = 0.9995. (b) $P = 20.6$ GPa ($T_\theta = 405 \pm 4$ K), fit quality R = 0.9992. (c) $P = 24.0$ GPa ($T_\theta = 398 \pm 4$ K), fit quality R = 0.9996. Yellow balls show $T_c$ defined by $R(T)/R_{norm}(T) = 0.95$ criterion; cyan balls show $T_c$ defined by $R(T)/R_{norm}(T) = 0.0$ criterion.

**Table 3.** Deduced $T_c$, $T_\theta$ and calculated $\lambda_{\text{e-ph}}$ for highly-compressed GeAs with assumed $\mu^* = 0.10$ [27].

| Pressure (GPa) | $T_\theta$ (K) | $T_c$ (K) | $\lambda_{e-ph,BCS\;asymp}$ | $\lambda_{e-ph,BCS}$ | $\lambda_{e-ph,aMcM}$ | $\lambda_{e-ph,fpc}$ [27] |
|---|---|---|---|---|---|---|
| 15.3 | 351 ± 5 | 7.3 $R(T)/R_{\text{norm}}(T) = 0.0$ | 0.041 ± 0.001 | 0.358 ± 0.001 | 0.612 ± 0.003 | $P = 10$ GPa 0.52 (Cm/2 phase) 0.65 (Rocksalt phase) |
| | | 8.1 $R(T)/R_{\text{norm}}(T) = 0.95$ | 0.042 ± 0.001 | 0.358 ± 0.001 | 0.634 ± 0.003 | |
| 20.6 | 405 ± 4 | 3.95 $R(T)/R_{\text{norm}}(T) = 0.0$ | 0.020 ± 0.001 | 0.316 ± 0.001 | 0.493 ± 0.001 | $P = 20$ GPa 0.51 (Rocksalt phase) |
| | | 6.3 $R(T)/R_{\text{norm}}(T) = 0.95$ | 0.031 ± 0.001 | 0.340 ± 0.001 | 0.558 ± 0.001 | |
| 24.0 | 398 ± 4 | 2.5 $R(T)/R_{\text{norm}}(T) = 0.0$ | 0.013(4) | 0.297 ± 0.001 | 0.446 ± 0.001 | $P = 20$ GPa 0.51 (Rocksalt phase) |
| | | 4.9 $R(T)/R_{\text{norm}}(T) = 0.95$ | 0.025(4) | 0.327 ± 0.001 | 0.523 ± 0.002 | |

### 4.4. Silane compressed at *P* = 192 GPa

First principles calculations for highly compressed hydrogen-rich silane, SiH$_4$, have been performed by several authors, from which we can mention a report by Feng *et al* [53] who computed $T_\theta \cong 3{,}500 - 4{,}000\;K$ and $T_c \cong 165\;K$, and the report by Chen *et al* [54] who computed $T_c = 20 - 75\;K$ for this hydrogen-rich material at pressure in the range of $70\;GPa \leq P \leq 250\;GPa$.

Eremets *et al* [28] discovered that SiH$_4$ exhibits low-temperature superconductivity with transition temperature in the range of $7\;K \leq T_c \leq 17\;K$ at pressure in the range of $60\;GPa \leq P \leq 192\;GPa$.

In Figure 8 we show raw $R(T)$ curve (from Fig. 2(b) of Ref. 28) and data fit to BG equation for SiH$_4$ compressed at $P = 192$ GPa. First of all, experimental data show that $T_\theta \cong 353 \pm 3\;K$, which is by one order of magnitude lower than $T_\theta \cong 3{,}500 - 4{,}000\;K$ computed by the first principles calculations [53].

To calculate $\lambda_{e-ph}$, we use $T_c$ = 7.7 K defined at the inflection point of temperature dependent resistance, $R(T=7.7$ K$)/R_{norm}(T)$ = 0.64, and $T_c$ = 10.6 K defined by $R(T=10.6$ K$)/R_{norm}(T)$ = 0.95 criterion. Computed values are in Table 4 for which μ* = 0.10 [55] and 0.12 [54] are used.

Taking in account that the ratio of $\frac{T_{c,0.64R_{norm}}}{T_\theta} = \frac{7.7\ K}{353\ K} \lesssim 0.022$, there is no need to consider intermediate or strong coupling scenarios. This becomes even more evident if we take in account that first principles calculations compute $\lambda_{e-ph,fpc} = 0.8 - 0.9$ which leads to unrealistically high $T_c$ values of $T_c = 20 - 165\ K$ [53,54].

**Table 4.** Deduced $T_\theta$ and calculated $\lambda_{e-ph}$ for highly-compressed silane at $P$ = 192 GPa.

| $T_\theta$ (K) | $T_c$ (K) | $\lambda_{e-ph,BCS\ asymp}$ | Assumed $\mu^*$ | $\lambda_{e-ph,BCS}$ | $\lambda_{e-ph,aMcM}$ | $\lambda_{e-ph,fpc}$ |
|---|---|---|---|---|---|---|
| 353 ± 3 | 7.7 $R(T)/R_{norm}(T)$ = 0.64 | 0.044 ± 0.001 | 0.10 [55] | 0.361 ± 0.001 | 0.622 ± 0.002 | 0.9 [55] |
| | | | 0.12 [54] | 0.381 ± 0.001 | 0.666 ± 0.002 | 0.8 [54] |
| | 10.6 $R(T)/R_{norm}(T)$ = 0.95 | 0.061 ± 0.001 | 0.10 [55] | 0.386 ± 0.001 | 0.699 ± 0.002 | 0.9 [55] |
| | | | 0.12 [54] | 0.406 ± 0.001 | 0.750 ± 0.002 | 0.8 [54] |

Thus, we can make an intermediate conclusion that at least for the first discovered highly-compressed hydrogen-rich compound, $SiH_4$, first principles calculations based on electron-phonon coupling mechanism and full $\alpha^2(\omega) \cdot F(\omega)$ spectrum are not able to reproduce, even approximately, observed in experiment $T_c$ (with the difference as large, as 20 times). This conclusion is also applied for highly-compressed black phosphorus, boron and GeAs.

The simplest assumption we can make is that in highly-compressed superconductors $T_c$ and $T_\theta$ are independent from each other.

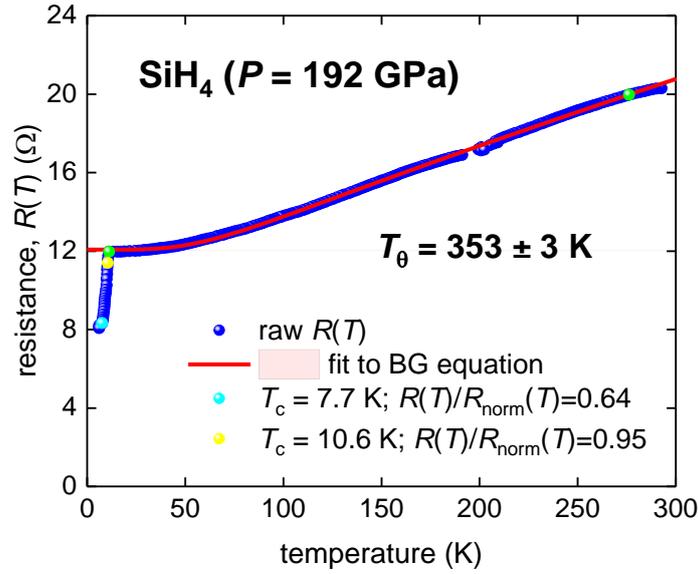

**Figure 8.** $R(T)$ data and fit to BG model (Eq. 5) for highly-compressed SiH$_4$ ($P$ = 192 GPa). Raw $R(T)$ data is from Ref. 28. 95% confidence bars are shown. Green balls indicate the bounds for which $\rho(T)$ data was used for the fit. Fit quality $R = 0.9994$. Cyan ball shows $T_c$ defined at the inflection point of $R(T)/R_{norm}(T) = 0.64$.

### 4.5. Highly compressed sulphur hydride

Drozdov *et al* [5] reported milestone discovery of NRT superconductivity in highly-compressed laser annealed sulphur hydride, H$_3$S, exhibited $T_c > 200\ K$ at megabar pressure.

Guigue *et al* [56] synthesized pure H$_3$S phase by laser heating hydrogen-embedded solid sulphur samples at pressures above 75 GPa. Diffraction studies showed that the compound has the crystal structure with space group of *Cccm* which exhibits up to pressure of $P = 160$ GPa. It should be noted, that *Cccm* phase of H$_3$S is non-superconducting. In contrast, Einaga *et al* [57] reported that H$_3$S compound synthesized from gaseous H$_2$S has low-pressure ($P \leq 150\ GPa$) low-$T_c$ phase with space group of *R3m*, and high-pressure ($P > 150\ GPa$) high-$T_c$ phase with space group of *Im-3m*.

Most extensive study for the phase transitions in highly-compressed sulphur hydride when gaseous H$_2$S is used as a precursor was reported by Goncharov *et al.* [58] who found a rich homological series of sulphur hydride phases, H$_n$S$_m$, which form at high-pressure and laser annealing conditions. Thus, phase composition/phase symmetry studies for highly-compressed sulphur hydride are at ongoing stage and the agreement between research groups be reached in a future. Here we report results on the evolution of $T_\theta$ and $\lambda_{e-ph,aMcM}$ vs applied pressure of $111\ GPa \leq P \leq 150\ GPa$ for low-$T_c$ *R3m*-superconducting phase of $H_3S$ compound reported by Einaga *et al* [57], as well as the analysis for four months ageing (at *P* = 155 GPa) and one freshly prepared sample at *P* = 160 GPa of high-$T_c$ *Im-3m* phase which were reported by Mozaffari *et al* [59].

### 4.5.1. *R3m* phase of $H_3S$

Einaga *et al.* [57] in their Fig. 3(a) reported *R*(*T*) curves for $H_3S$ measured at applied pressure in the range of $111\ GPa \leq P \leq 150\ GPa$. All reported *R*(*T*) curves reach *R* = 0 Ω point, however samples subjected to pressures of *P* = 133 GPa and 150 GPa have long tails to reach zero resistance with inflection points at $R(T)/R_{norm}(T) \sim 0.05$, where $R_{norm}(T)$ is extrapolated curve of *R*(*T*) fit to BG equation (Eq. 5). Based on this, $T_c$ is defined by $R(T)/R_{norm}(T) = 0.05$ criterion for compressed sulphur hydride samples considered in this Section.

At pressure range of $111\ GPa \leq P \leq 150\ GPa$ annealed $H_3S$ compound exhibits in *R3m* phase. Taking in account that experiment [57] shows that $T_c$ is monotonically changing vs applied pressure there is an expectation that $T_\theta$ and $\lambda_{e-ph}$ will be also following monotonic trends. However, the analysis of experimental *R*(*T*) data (Fig. 9, Table 5) shows that the $T_\theta$ and $\lambda_{e-ph}$ are varying in a random way in comparison with $T_c$ (Fig. 10), which is an evidence that in *R3m* phase of $H_3S$, one of integrated characteristic of the phonon spectrum, which is $T_\theta$, does not correlate with the superconducting transition temperature, $T_c$.

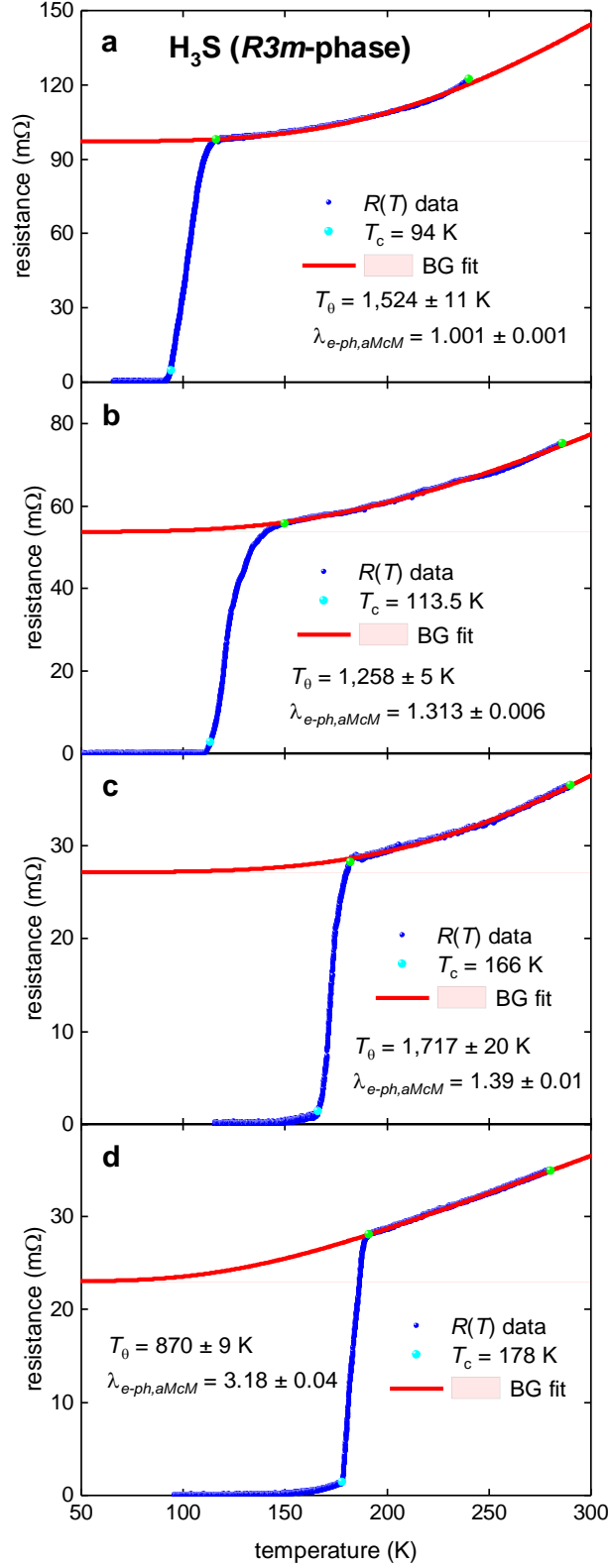

**Figure 9.** $R(T)$ data and fit to BG model for highly-compressed *R3m*-phase of $H_3S$ ($111\ GPa \leq P \leq 150\ GPa$). Raw $R(T)$ data is from Ref. [31]. 95% confidence bars are shown. Green balls indicate the bounds for which $R(T)$ data was used for the fit. Cyan ball shows $T_c$ defined by the $R(T)/R_{norm}(T) = 0.05$ criterion. Showed $\lambda_{e-ph,aMcM}$ values calculated for $\mu^* = 0.10$. (a) Fit quality $R = 0.997$; (b) $R = 0.998$; (c) $R = 0.998$; (d) $R = 0.9993$.

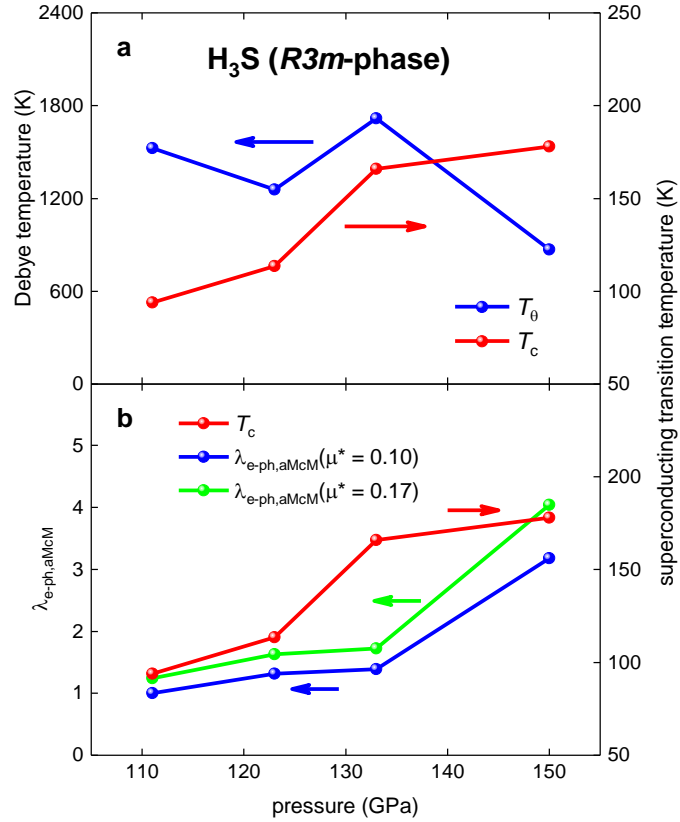

**Figure 10.** (a) Deduced Debye temperatures, $T_\theta$, and superconducting transition temperature, $T_c$, for highly-compressed *R3m*-phase of $H_3S$ ($111\ GPa \leq P \leq 150\ GPa$) vs applied pressure. (b) Superconducting transition temperature, $T_c$, and calculated electron-phonon coupling strength constant for highly-compressed *R3m*-phase of $H_3S$ ($111\ GPa \leq P \leq 150\ GPa$) vs applied pressure. Raw data is from Ref. 57.

**Table 5.** Deduced $T_\theta$ and calculated $\lambda_{e-ph}$ for highly-compressed *R3m*-phase of $H_3S$ at $111\ GPa \leq P \leq 150\ GPa$. $T_c$ values are defined by $R(T)/R_{norm}(T) = 0.05$ criterion.

| Pressure (GPa) | $T_\theta$ (K) | $T_c$ (K) | Assumed $\mu^*$ | $\lambda_{e-ph,BCS}$ | $\lambda_{e-ph,aMcM}$ | $\lambda_{e-ph}$ (first-principles calculations) [60] |
|---|---|---|---|---|---|---|
| 111 | 1524 ± 11 | 94.0 | 0.10 | 0.459 ± 0.001 | 1.001 ± 0.004 | |
| | | | 0.17 | 0.529 ± 0.001 | 1.241 ± 0.006 | |
| 123 | 1258 ± 5 | 113.5 | 0.10 | 0.516 ± 0.001 | 1.313 ± 0.006 | |
| | | | 0.17 | 0.586 ± 0.001 | 1.628 ± 0.006 | 2.07 |
| 133 | 1717 ± 20 | 166.0 | 0.10 | 0.528 ± 0.002 | 1.391 ± 0.013 | |
| | | | 0.17 | 0.598 ± 0.002 | 1.726 ± 0.017 | |
| 150 | 870 ± 9 | 178.0 | 0.10 | 0.730 ± 0.002 | 3.18 ± 0.04 | |
| | | | 0.17 | 0.805 ± 0.003 | 4.04 ± 0.06 | |

It should be noted that deduced $\lambda_{e-ph,aMcM}$ values (Table 5) cover so wide range of $1.00 \leq \lambda_{e-ph,aMcM} \leq 4.03$ that there is no possibility to affirm/disprove theoretical value of $\lambda_{e-ph,aMcM} = 2.07$ reported for *R3m*-phase by Duan *et al* [60]. However, there is another way to utilize Eqs. 5,11 because ones do not only apply for two isotopic counterparts, but also can be apply for the same compound at the same phase state when the superconducting transition temperature of the material is changing vs the change in the pressure:

$$\left.\frac{T_{c,n}}{T_{c,m}}\right|_{exp} = \left.\frac{T_{\theta,n}}{T_{\theta,m}}\right|_{exp}, \tag{28}$$

where the subscript *m* indicates a *m*-stage of the compression and *n* indicates *n*-stage of the compression. If the NRT superconductivity is mediated by the electron-phonon interaction, then Eq. 5 should be valid. In Fig. 11 we show data for *R3m*-phase of $H_3S$ for ratios of:

$$\left.\frac{T_{c,n}}{T_{c,P=111\ GPa}}\right|_{exp} = \left.\frac{T_{\theta,n}}{T_{\theta,P=111\ GPa}}\right|_{exp}, \tag{29}$$

Results (Fig. 3) are in a large disagreement with the assumption that the NRT superconductivity is mediated by the electron-phonon mechanism in *R3m*-phase of $H_3S$. The most compelling case is for pressure of *P* = 150 GPa, where the disagreement between expected and deduced (from experiment) values is in more than three times.

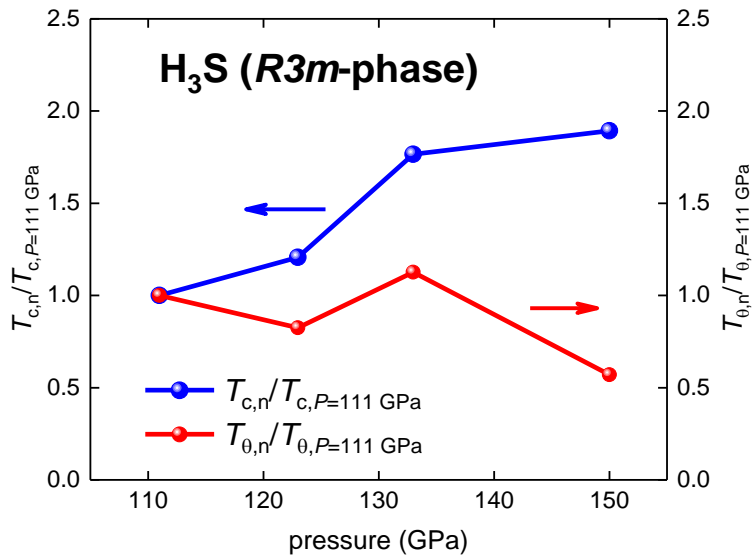

**Figure 11.** The ratios of $\frac{T_{c,n}}{T_{c,P=111\ GPa}}$ and $\frac{T_{\theta,n}}{T_{\theta,P=111\ GPa}}$ for highly-compressed *R3m*-phase of $H_3S$ ($111\ GPa \leq P \leq 150\ GPa$) vs applied pressure.

### 4.5.2. *Im-3m* phase of H₃S

The first H₃S sample was subjected to $P$ = 155 GPa. $R(T)$ curves measured at different ageing stages (within 4 months) show two superconducting transitions in the temperature range of $192.5\ K < T_c < 201\ K$ (Fig. 12), however, all $R(T)$ curves for this sample reach zero resistance point, $R(T) = 0\ \Omega$, at $T > 192.5\ K$.

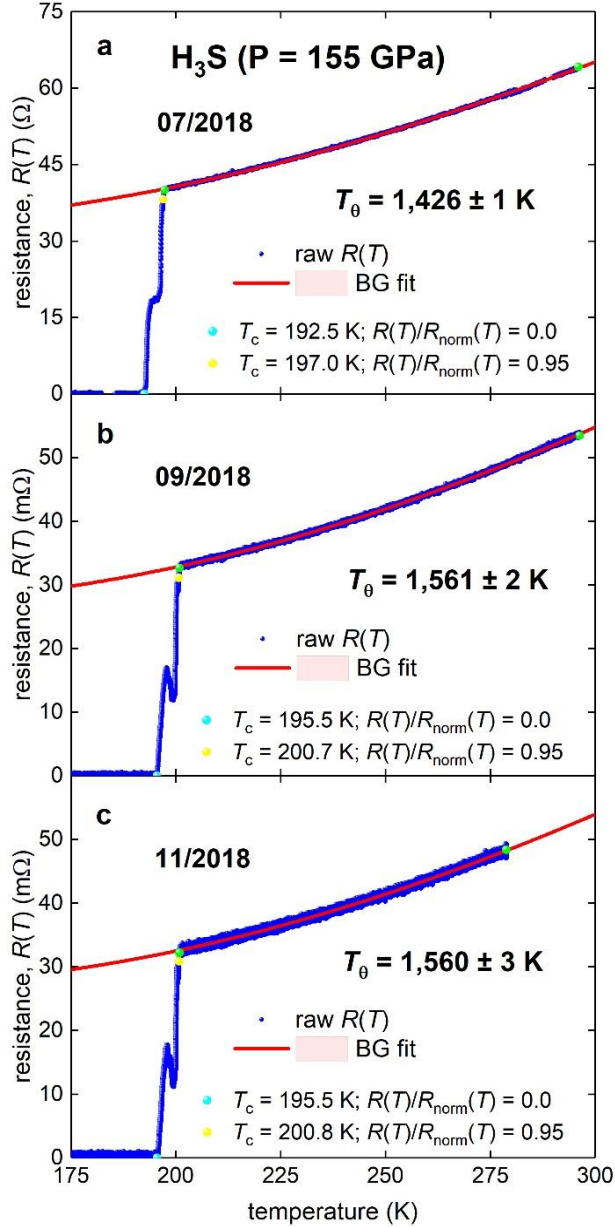

**Figure 12.** $R(T)$ data and fit to BG model (Eq. 5) for highly-compressed *Im-3m* phase of H₃S ($P$ = 155 GPa) at different ageing stages. Raw $R(T)$ data is from Ref. 59. 95% confidence bars are shown. Green balls indicate the bounds for which $\rho(T)$ data was used for the fit. Cyan balls show $T_c$ defined by $R(T)/R_{norm}(T) = 0.0$ criterion. Yellow balls show $T_c$ defined by $R(T)/R_{norm}(T) = 0.95$ criterion. (a) Fit quality $R = 0.9998$. (b) Fit quality $R = 0.9995$. (c) Fit quality $R = 0.9971$.

The second sample subjected to pressure of $P = 160$ GPa has $R(T)$ curve which does not reach zero resistance point (Fig. 10) even at lowest available in experiment temperature of $T = 82.9$ K. For this sample, $T_c$ is defined by the inflection point of $R(T=158$ K$)/R_{norm}(T) = 0.08$, while $R(T=183$ K$)/R_{norm}(T) = 0.95$ (Fig. 13).

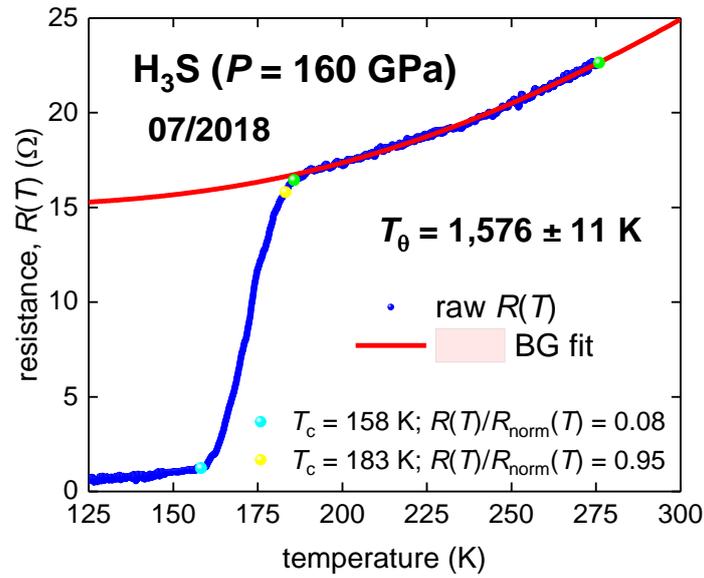

**Figure 13.** $R(T)$ data and fit to BG model (Eq. 5) for highly-compressed $H_3S$ ($P = 160$ GPa). Raw $R(T)$ data is from Ref. 59. 95% confidence bars are shown. Green balls indicate the bounds for which $\rho(T)$ data was used for the fit. Cyan ball shows $T_c$ defined by the inflection point of $R(T)/R_{norm}(T) = 0.08$. Yellow ball shows $T_c$ defined by $R(T)/R_{norm}(T) = 0.95$ criterion. Fit quality $R = 0.997$.

Deduced $T_\theta$ and computed $\lambda_{e-ph}$ for both samples are shown in Table 6. First of all, we should mention remarkable constancy in deduced Debye temperature $T_\theta = 1531 \pm 70$ K for both samples and different ageing stages for the first sample, with standard deviation less than 2%.

It can be also seen in Table 6, that as $T_c$, as $\lambda_{e-ph}$ are varied in wide ranges covers values computed for harmonic and anharmonic models, when different criteria and coupling-strength models are applied, and thus there are no experimental evidences to make a conclusion which model is more preferable in comparison with others.

However, deduced from experiment the Debye temperature, $T_\theta$, is remaining to be constant for all samples. This result is in agreement with the report by Harshman and Fiory [61] who performed analyses of residual resistance ratios for $H_3S$ and $D_3S$ and found that $T_c$ values and the transition width of the transition in $H_3S$-$D_3S$ system are related to the samples purity and atomic disorder, which overwhelming the influence on $T_c$ originated by other factors.

**Table 6.** Deduced $T_\theta$ and calculated $\lambda_{e\text{-ph}}$ for highly-compressed $H_3S$ at $P = 155$ GPa (raw $R(T)$ curves were reported by Mozaffari *et al* [59]).

| Sample ID | $T_\theta$ (K) | $T_c$ (K) | $\lambda_{e-ph,asymp}$ | Assumed $\mu^*$ | $\lambda_{e-ph,BCS}$ | $\lambda_{e-ph,aMcM}$ | $\lambda_{e-ph}$ (first-principles calculations) [37] |
|---|---|---|---|---|---|---|---|
| 07/2018 ($P = 155$ GPa) | $1426 \pm 1$ | 192.5 $R(T)/R_{norm}(T) = 0.0$ | 0.270 | 0.10 | 0.599 | 1.93 | |
| | | | | 0.15 | 0.649 | 2.26 | |
| | | 197.0 $R(T)/R_{norm}(T) = 0.95$ | 0.276 | 0.10 | 0.605 | 1.98 | |
| | | | | 0.15 | 0.655 | **2.33** | |
| 09/2018 ($P = 155$ GPa) | $1561 \pm 2$ | 195.5 $R(T)/R_{norm}(T) = 0.0$ | 0.251 | 0.10 | 0.582 | 1.78 | 1.84 (anharmonic) 2.64 (harmonic) |
| | | | | 0.15 | 0.632 | 2.09 | |
| | | 200.7 $R(T)/R_{norm}(T) = 0.95$ | 0.257 | 0.10 | 0.588 | 1.83 | |
| | | | | 0.15 | 0.638 | 2.15 | |
| 11/2018 ($P = 155$ GPa) | $1560 \pm 3$ | 195.5 $R(T)/R_{norm}(T) = 0.0$ | 0.251 | 0.10 | 0.582 | 1.78 | |
| | | | | 0.15 | 0.632 | 2.09 | |
| | | 200.8 $R(T)/R_{norm}(T) = 0.95$ | 0.257 | 0.10 | 0.588 | 1.83 | |
| | | | | 0.15 | 0.638 | 2.15 | |
| 07/2018 ($P = 160$ GPa) | $1576 \pm 11$ | 158 $R(T)/R_{norm}(T) = 0.08$ | 0.201 | 0.10 | 0.535 | **1.44** | |
| | | | | 0.15 | 0.585 | 1.68 | |
| | | 183 $R(T)/R_{norm}(T) = 0.95$ | 0.232 | 0.10 | 0.564 | 1.65 | |
| | | | | 0.15 | 0.614 | 2.15 | |

## 4.6. Highly compressed D$_3$S

Drozdov et al [5] in their Fig. 2(b) reported $R(T)$ dataset for highly-compressed D$_3$S ($P$ = 155 GPa). $R(T)$ data fit to Eq. 5 is shown in Fig. 14. Reasonably large uncertainty in deduced $T_\theta = 982 \pm 127\ K$ is primarily related to narrow temperature range for which $R(T)$ data was measured. We should mention that sulphur deuteride at $P$ = 155 GPa exhibits *R3m* phase [57], which is different from *Im-3m* phase of H$_3$S compound at the same pressure.

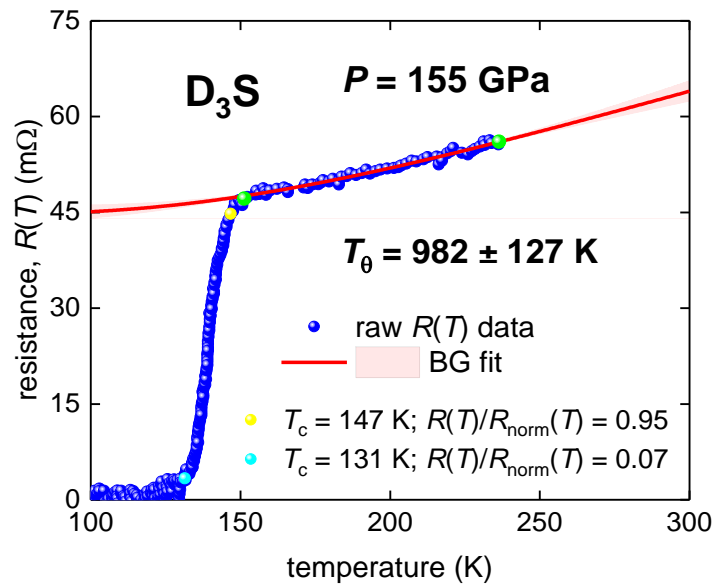

**Figure 14.** $R(T)$ data and fit to BG model (Eq. 5) for highly-compressed D$_3$S in *R3m* phase ($P = 155\ GPa$). Raw $R(T)$ data is from Ref. 5. 95% confidence bars are shown. Green balls indicate the bounds for which $\rho(T)$ data was used for the fit. Cyan ball shows $T_c$ defined by $R(T)/R_{norm}(T) = 0.07$ criterion. Yellow ball shows $T_c$ defined by $R(T)/R_{norm}(T) = 0.95$ criterion. Fit quality $R = 0.978$.

In Figs. 15,16 we show $R(T)$ data fits for D$_3$S ($P$ = 173 and 190 GPa, respectively) reported by Einaga et al [57] in their Fig. 3(b). Calculated $\lambda_{e\text{-ph}}$ for all three D$_3$S samples are in Table 7.

It should be noted that due to D$_3$S samples were subjected to a wide range of pressure, $P = 155 - 190\ GPa$, at lowest pressure the compounds has *R3m* phase symmetry, and for $P = 173\ and\ 190\ GPa$ the compound has *Im-3m* phase symmetry (which is exhibited at $P \geq 160\ GPa$).

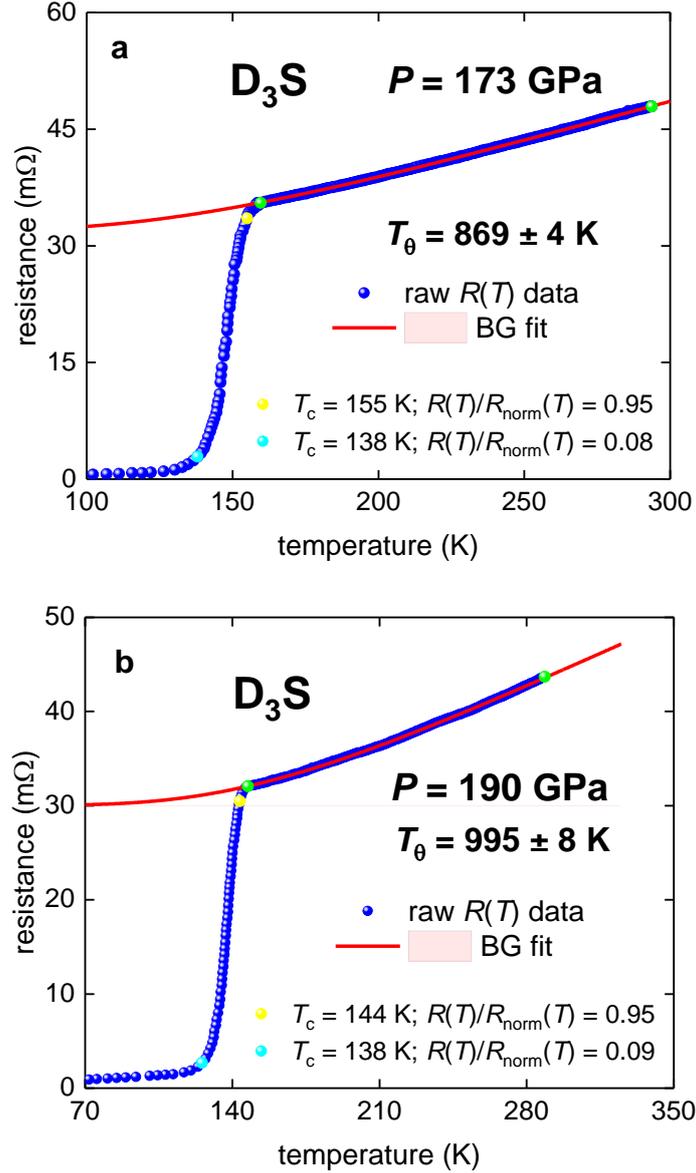

**Figure 15.** $R(T)$ data and fit to BG model (Eq. 5) for highly-compressed $D_3S$ in *Im-3m* phase. Raw $R(T)$ data is from Ref. 58. 95% confidence bars are shown. Green balls indicate the bounds for which $\rho(T)$ data was used for the fit. (a) $P = 173\ GPa$. Cyan ball shows $T_c$ defined by $R(T)/R_{norm}(T) = 0.08$ criterion. Yellow ball shows $T_c$ defined by $R(T)/R_{norm}(T) = 0.95$ criterion. Fit quality $R = 0.99992$. (b) $P = 173\ GPa$. Cyan ball shows $T_c$ defined by $R(T)/R_{norm}(T) = 0.09$ criterion. Yellow ball shows $T_c$ defined by $R(T)/R_{norm}(T) = 0.95$ criterion. Fit quality $R = 0.9998$.

Based on this, we calculate average $T_\theta$ only for *Im-3m* phase:

$$T_\theta = 930 \pm 92\ K \tag{30}$$

which has, nevertheless, a narrow uncertainty of 10%. However, due to its large uncertainty, the $T_\theta$ value for *R3m* phase is also within this temperature range.

In overall, deduced $\lambda_{e-ph}$ (Table 7) for *Im-3m* phase are equally well matched as harmonic, as anharmonic models, because the highest deduced $\lambda_{e-ph}$ value is agreed with the former, and the lowest $\lambda_{e-ph}$ value is agreed with the latter (the highest and the lowest $\lambda_{e-ph}$ values are marked in bold in Table 7).

**Table 7.** Deduced $T_\theta$ and calculated $\lambda_{e-ph}$ for D$_3$S compressed at $P = 155, 173, and\ 190\ GPa$ (raw $R(T)$ curves were reported by Drozdov *et al* [5] and Einaga *et al* [57]).

| Pressure (GPa) | $T_\theta$ (K) | $T_c$ (K) | $\lambda_{e-ph,asymp}$ | Assumed $\mu^*$ | $\lambda_{e-ph,BCS}$ | $\lambda_{e-ph,aMcM}$ | $\lambda_{e-ph}$ (first-principles calculations) |
|---|---|---|---|---|---|---|---|
| 155 (Drozdov *et al* [5]) | 982 ± 127 | 131 $R(T)/R_{norm}(T) = 0.07$ | 0.267 ± 0.031 | 0.10 | 0.597 ± 0.029 | 1.91 ± 0.23 | *R3m* phase |
| | | | | 0.15 | 0.647 ± 0.029 | 2.24 ± 0.29 | |
| | | 147 $R(T)/R_{norm}(T) = 0.95$ | 0.299 ± 0.031 | 0.10 | 0.627 ± 0.032 | 2.17 ± 0.28 | |
| | | | | 0.15 | 0.677 ± 0.032 | 2.55 ± 0.33 | |
| 173 (Einaga *et al* [58]) | 869 ± 4 | 138 $R(T)/R_{norm}(T) = 0.08$ | 0.318 ± 0.002 | 0.10 | 0.644 ± 0.002 | 2.33 ± 0.01 | *Im-3m* phase 1.87 (anharmonic) [37] 2.64 (harmonic) [37] |
| | | | | 0.15 | 0.694 ± 0.002 | **2.74 ± 0.01** | |
| | | 155 $R(T)/R_{norm}(T) = 0.95$ | 0.357 ± 0.002 | 0.10 | 0.680 ± 0.002 | 2.68 ± 0.02 | |
| | | | | 0.15 | 0.730 ± 0.002 | 3.17 ± 0.02 | |
| 190 (Einaga *et al* [58]) | 995 ± 8 | 138 $R(T)/R_{norm}(T) = 0.09$ | 0.275 ± 0.003 | 0.10 | 0.606 ± 0.002 | **1.99 ± 0.02** | |
| | | | | 0.15 | 0.656 ± 0.002 | 2.34 ± 0.02 | |
| | | 144 $R(T)/R_{norm}(T) = 0.95$ | 0.290 ± 0.003 | 0.10 | 0.617 ± 0.002 | 2.09 ± 0.02 | |
| | | | | 0.15 | 0.667 ± 0.002 | 2.45 ± 0.02 | |

It should be also noted, that, in overall, $\lambda_{e-ph}$ values deduced for D$_3$S are much higher than $\lambda_{e-ph}$ deduced for H$_3$S, which is in remarkable contradiction with results of first-principles calculations, which showed that $\lambda_{e-ph}$ for both isotopic counterparts are indistinguishably close to each other [37]. More detailed discussion of the isotope effect in H$_3$S-D$_3$S system is given in next Section.

### 4.7. Isotope effect in H₃S-D₃S system

Now we turn to the discussion of the most crucial effect which can confirm or disprove the electron-phonon coupling mechanism in H₃S-D₃S system, which is the isotope effect. Due to all three available raw $R(T)$ datasets for D₃S never reached $R(T)/R_{norm}(T) = 0.0$ state, we compare herein $T_c$ data for H₃S and D₃S for $R(T)/R_{norm}(T) = 0.95$ criterion.

As we show above that if the superconductivity in two these isotope counterparts is belonging the electron-phonon coupling, then the Eq. 16 should be satisfied. By taking in account primary conclusion of first-principles calculations [37], that H₃S and D₃S are anharmonic electron-phonon superconductors:

$$\left.\frac{\hbar \cdot \omega_{ln,H3S}}{\hbar \cdot \omega_{ln,D3S}}\right|_{fpc,anharmonic} = \left.\frac{92.86 \, meV}{73.3 \, meV}\right|_{fpc,anharmonic} = 1.27 \qquad (31)$$

The ratio in Eq. 31 should be compare with the ratio of average value for $T_c$ for two isotopic counterparts in *Im-3m* crystallographic phase state (Tables 6 and 7):

$$\left.\frac{T_{c,H3S}}{T_{c,D3S}}\right|_{R(T)/R_{norm}(T)=0.95} = \left.\frac{195.4 \pm 8.4 \, K}{149.5 \pm 7.8 \, K}\right|_{R(T)/R_{norm}(T)=0.95} = 1.31 \pm 0.05 \qquad (32)$$

which is in a good agreement with Eq. 31.

The ratio of the Debye temperatures for two isotopic counterparts in *Im-3m* crystallographic phase (Tables 6 and 7) is:

$$\frac{T_{\theta,H3S}}{T_{\theta,D3S}} = \frac{1531 \pm 70 \, K}{930 \pm 92 \, K} = 1.65 \pm 0.17 \qquad (33)$$

### 4.8. Superconductors in LaH$_x$-LaD$_y$ system

#### 4.8.1. LaH₁₀ with $T_c$ = 240 K

We start our analysis of LaH$_x$-LaD$_y$ system by the analysis of $R(T)$ dataset for Sample 3 which exhibits the highest transition temperature of $T_c$ = 240 K in the report by Drozdov *et al* [62] in their Fig. 16. This sample has two inflection points in $R(T)$ curves which can be seen in in Fig. 16, and we calculate $\lambda_{e-ph,aMcM}$ values for these two superconducting transition temperatures (Table 8). In our calculations we use μ* = 0.10 reported by Errea *et al.* [63].

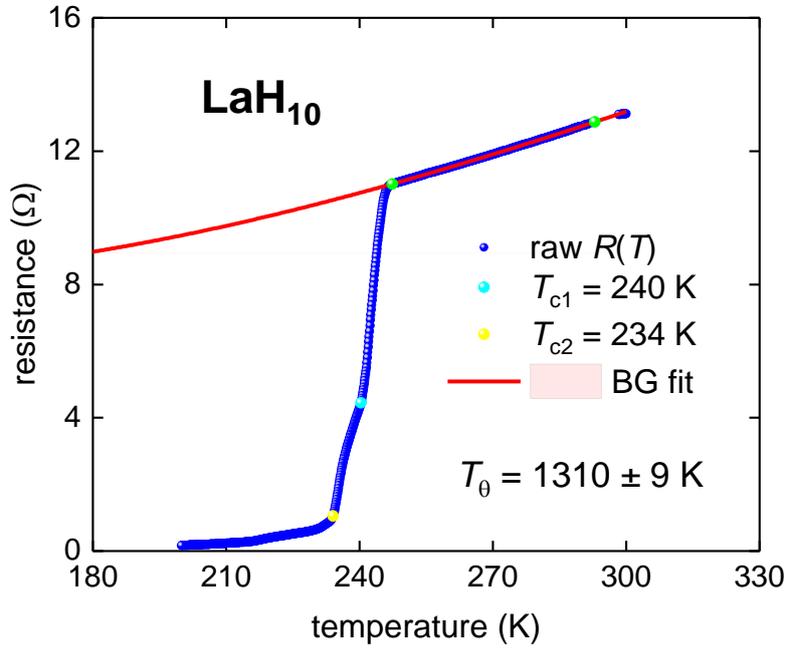

**Figure 16.** $R(T)$ data and fit to BG model for highly-compressed LaH$_{10}$ (Sample 3 [62]). Raw $R(T)$ data is from Ref. [62]. 95% confidence bars are shown. Green balls indicate the bounds for which $R(T)$ data was used for the fit. Cyan and yellow balls show $T_c$ defined by two inflection points. Fit quality $R = 0.99992$.

**Table 8.** Deduced $T_\theta$ and calculated $\lambda_{e\text{-ph,aMcM}}$ for highly-compressed LaH$_{10}$ (Sample 3 [62]). Assumed μ* = 0.10 [63]. $T_c$ values defined by the inflection points in $R(T)$ curve.

| Compound (Sample ID; pressure) | $T_\theta$ (K) | $T_c$ (K) | $\lambda_{e-ph,BCS}$ | $\lambda_{e-ph,aMcM}$ | $\lambda_{e-ph}$ (first-principles calculations) [63] |
|---|---|---|---|---|---|
| LaH$_{10}$ (Sample 3; $P$ = 150 GPa) | 1310 ± 9 | 240 | 0.689 ± 0.002 | 2.77 ± 0.02 | 2.76 ($P$ = 163 GPa) |
|  |  | 234 | 0.681 ± 0.002 | 2.69 ± 0.02 |  |

One can see an excellent agreement between computed (by first principles calculations [63]) and deduced (by our analysis herein) the electron-phonon coupling strength values, $\lambda_{e-ph}$, for LaH$_{10}$ compound (Table 8). However, the rest of available experimental datasets shows the large disagreement between computed and deduced (from experiment) values, which we present below. And these majority of disagreement case, as we mentioned above, should be considered with equal weight with successful cases, to understand the nature of NRT superconductivity in highly-compressed super-hydrides/deuterides.

### 4.8.2. LaH$_x$ with $T_c$ ~ 215 K

Drozdov *et al* [62] in their Extended Data Fig. 5 reported $R(T)/R_{norm}(T)$ curve for laser annealed LaH$_x$ (Sample 12) with very sharp superconducting transition with $T_c$ ~ 210 K at $P$ = 160 GPa. When the pressure was decreased to $P$ = 150 GPa, the transition temperature increased to $T_c$ ~ 215 K. Reduced resistance curve, $R(T)/R_{norm}(T)$, at $P$ = 150 GPa is analysed in Fig. 17 with deduced parameters show in Table 9.

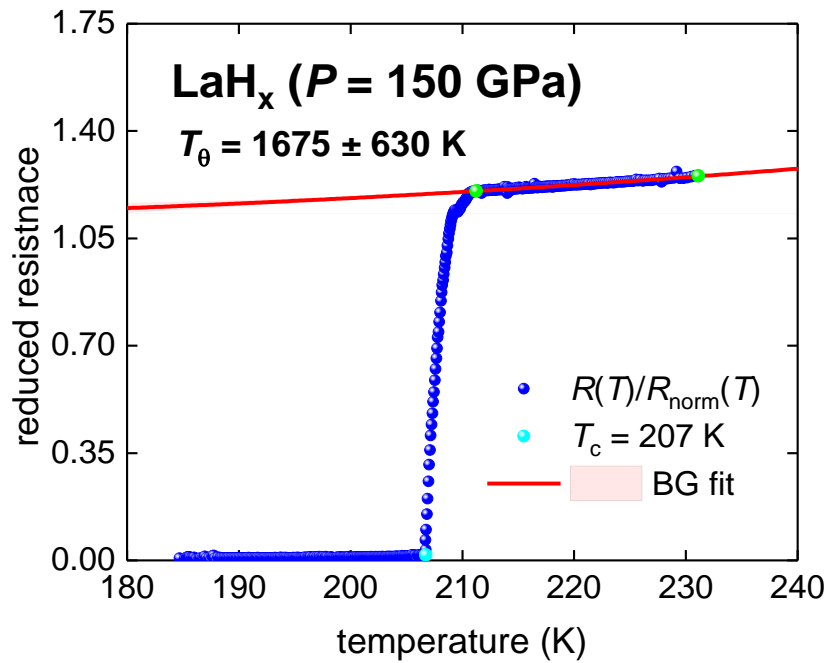

**Figure 17.** $R(T)/R_{norm}(T)$ data and fit to BG model for highly-compressed LaH$_{10}$ (Sample 12 [62]). 95% confidence bars are shown. Green balls indicate the bounds for which $R(T)$ data was used for the fit. Cyan ball shows $T_c$ defined by the $R(T)/R_{norm}(T) = 0.01$ criterion. Fit quality $R = 0.959$.

**Table 9.** Deduced $T_\theta$ and calculated $\lambda_{e\text{-}ph,aMcM}$ for highly-compressed LaH$_x$ (Sample 12 [62]). Assumed $\mu^* = 0.10$ [63]. $T_c$ value defined by zero resistance point.

| Compound (Sample ID; pressure) | $T_\theta$ (K) | $T_c$ (K) | $\lambda_{e-ph,BCS}$ | $\lambda_{e-ph,aMcM}$ | $\lambda_{e-ph}$ (first-principles calculations) [34] |
|---|---|---|---|---|---|
| LaH$_x$ (Sample 12; $P$ = 150 GPa) | 1675 ± 630 | 207 | 0.58 ± 0.14 | $1.75^{+1.4}_{-0.4}$ | 2.67-3.62 |

Due to the normal part of the resistance curve for this sample is relatively narrow, the uncertainty in deduced Debye temperature is large, however, calculated value of $\lambda_{e-ph,aMcM} = 1.75^{+1.4}_{-0.4}$ seems to be still far apart from computed value range of $\lambda_{e-ph,aMcM} = 2.67 - 3.62$ reported by Errea et al [63].

### 4.8.3. LaH$_x$ and LaD$_y$ with $T_c \sim 65$ K

Drozdov et al [62] in their Extended Data Fig. 5 reported $R(T)/R_{norm}(T)$ curves for LaH$_x$ (Sample 11) and LaD$_y$ (Sample 14) with very close superconducting transition temperatures. By use of the $R(T)/R_{norm}(T) = 0.05$ criterion, the transition temperatures are found to be $T_c = 66.2$ K and $T_c = 65.3$ K for LaH$_x$ and LaD$_y$ respectively. This is practically ideal pair to test the validity of electron-phonon mediated NRT superconductivity in LaH-LaD system, because the ratio of transition temperatures for these isotopic counterparts is practically undistinguishable from the unity:

$$\left.\frac{T_{c,2}}{T_{c,1}}\right|_{exp} = \frac{66.2\ K}{65.3\ K} = 1.014 \cong 1.0 \tag{34}$$

where subscripts 1 and 2 designate LaH$_x$ and LaD$_y$ compounds respectively.

In Fig. 18 and Table 10 we show temperature dependent of the reduced resistance and data fits to BG model. Deduced ratio for the Debye temperatures for these isotopic counterparts is:

$$\left.\frac{T_{\theta,2}}{T_{\theta,1}}\right|_{exp} = \frac{603\ K}{415\ K} = 1.453 \cong 1.5 \tag{35}$$

which is remarkably different from the ratio of the transition temperatures (Eq. 17). From this we can conclude there is no experimental evidences that electron-phonon mechanism is the origin for superconductivity in these low-$T_c$ samples of LaH$_x$ and LaD$_y$.

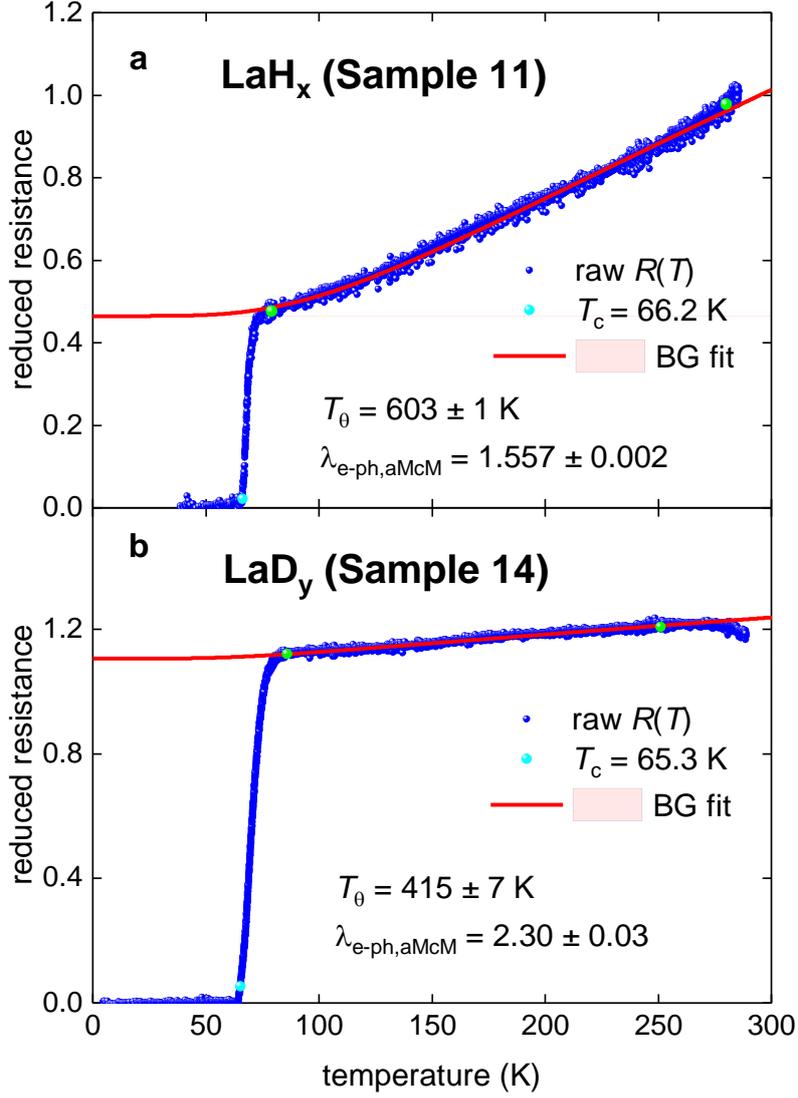

**Figure 18.** $R(T)$ data and fit to BG model for highly-compressed $LaH_{10}$ (Sample 11 [62], panel **a**) and $LaD_{10}$ (Sample 14 [62], panel **b**). Raw $R(T)$ data is from Ref. [62]. 95% confidence bars are shown. Green balls indicate the bounds for which $R(T)$ data was used for the fit. Cyan ball shows $T_c$ defined by the $R(T)/R_{norm}(T) = 0.05$ criterion. (a) Fit quality $R = 0.997$; (b) $R = 0.982$.

**Table 10.** Deduced $T_\theta$ and calculated $\lambda_{e\text{-ph,aMcM}}$ for highly-compressed $LaH_x$ (Sample 11 [62]) and $LaD_y$ (Sample 14 [62]). Assumed $\mu^* = 0.10$ [63]. $T_c$ values defined by the inflection points in $R(T)$ curve. $T_c$ values are defined by $R(T)/R_{norm}(T) = 0.05$ criterion.

| Compound (Sample ID; pressure) | $T_\theta$ (K) | $T_c$ (K) | $\lambda_{e-ph,BCS}$ | $\lambda_{e-ph,aMcM}$ | $\lambda_{e-ph}$ (first-principles calculations) [34] |
|---|---|---|---|---|---|
| $LaH_x$ (Sample 11; $P = 150$ GPa) | $603 \pm 1$ | 66.2 | $0.552 \pm 0.001$ | $1.557 \pm 0.002$ | 2.67-3.62 |
| $LaD_y$ (Sample 14; $P = 130$ GPa) | $415 \pm 7$ | 65.3 | $0.641 \pm 0.002$ | $2.30 \pm 0.03$ | 3.14 |

### 4.8.4. LaH$_x$ and LaD$_{11}$ with $T_c$ ~ 100 K

Drozdov *et al* [62] in their Extended Data Fig. 5 reported $R(T)/R_{norm}(T)$ curves for different laser annealing stage of LaH$_x$ (Sample 10) and LaD$_{11}$ (Sample 8) specimens which have superconducting transition temperatures near 100 K, if the transition will be defined by the inflection point (for Samples in Fig. 19,a and 19,c) or by the for criterion of $R(T)/R_{norm}(T)$ = 0.25 (Fig. 19,b). Taking in account that $R(T)/R_{norm}(T)$ curve for LaD$_{11}$ (Sample 8, Fig. 19,b) has broad low-temperature tail with clearly observed inflection point at $T$ = 125 K for which the $T_c$ criterion is $R(T)/R_{norm}(T)$ = 0.25, the same $T_c$ criterion was applied for laser annealed LaH$_x$ counterpart (Sample 10, Fig. 19,b), for which the transition temperature defines as $T_c$ = 107 K (Fig. 19,b).

Thus, in Table 6 we calculated $\lambda_{e-ph,aMcM}$ values for these three compounds with deduced $T_\theta$ from the fit of $R(T)/R_{norm}(T)$ curve to BG equation which are shown in Fig. 19.

Laser-annealed isotopic counterparts LaH$_x$ (Sample 10, Fig. 19,b) and LaD$_{11}$ (Sample 8) have reasonably close ratios:

$$\left.\frac{T_{c,2}}{T_{c,1}}\right|_{exp} = \frac{125\,K}{107\,K} = 1.17 \neq \left.\frac{T_{\theta,1}}{T_{\theta,2}}\right|_{exp} = \frac{1199\,K}{941\,K} = 1.27 \tag{36}$$

where subscripts 1 and 2 designate LaH$_x$ (Sample 10, Fig. 19,b) and LaD$_{11}$ compounds respectively. It should be noted, that for this NRT pair, $T_c$ and $T_\theta$ for hydrogen-based compound (i.e. LaH$_x$) are lower than ones for deuterium-based compound. In overall, computed $\lambda_{e-ph}$ values by Errea *et al.* [63] in assumption of $\mu^*$ = 0.10 (Table 11) for these NRT superconductors are very different from deduced $\lambda_{e-ph}$ value we deduced in our analysis herein (Table 11).

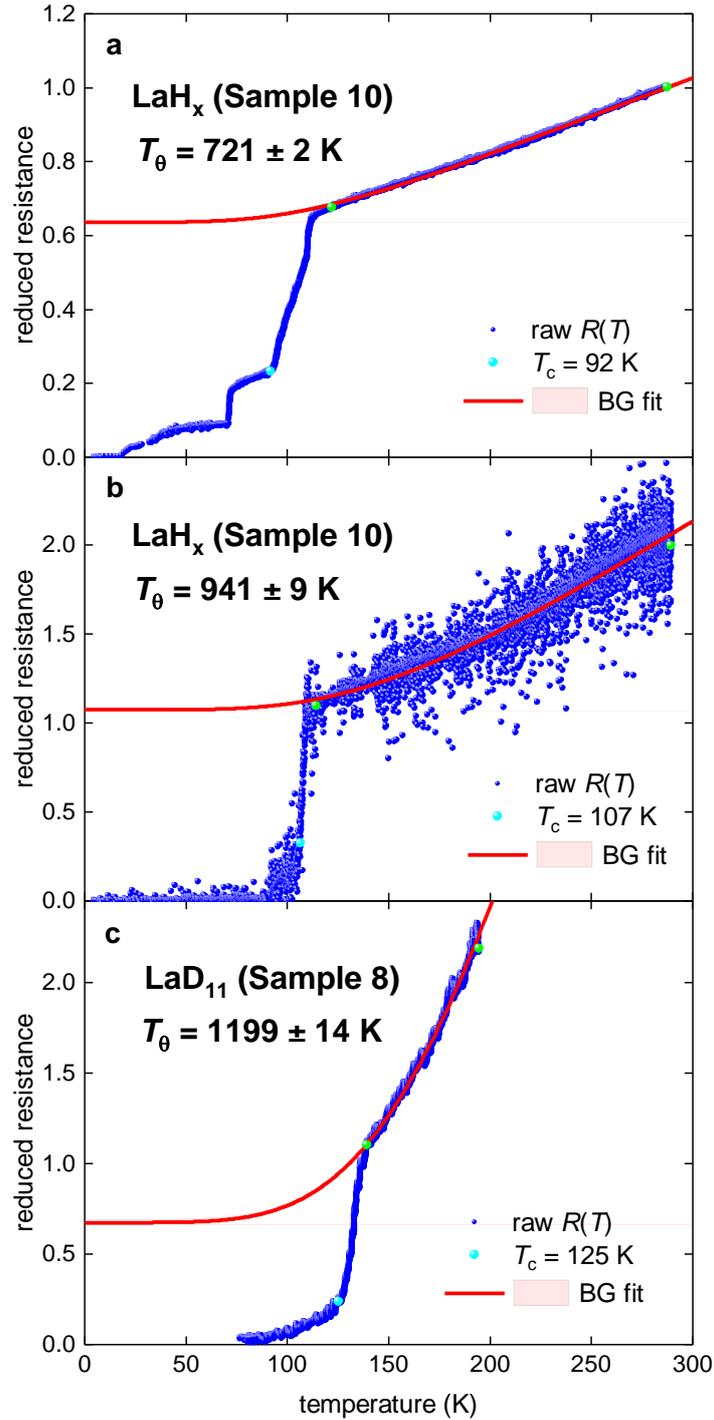

**Figure 19.** Temperature dependent reduced resistance data and fit to BG model for highly-compressed LaH$_x$ (Sample 10 [62], panels **a,b**) and LaD$_{11}$ (Sample 8 [62], panel **c**). LaH$_x$ samples in panels **a** and **b** are at different laser annealing stages. Raw data is from Ref. [62]. 95% confidence bars are shown. Green balls indicate the bounds for which resistance data is used for the fit. Cyan ball shows $T_c$ defined by criteria described in main text. (a) Fit quality $R = 0.9990$; (b) $R = 0.944$; (c) $R = 0.992$.

**Table 11.** Deduced $T_\theta$ and calculated $\lambda_{\text{e-ph,aMcM}}$ for highly-compressed $\text{LaH}_x$ (Sample 10 [62]) and $\text{LaD}_{11}$ (Sample 8 [62]). Assumed $\mu^* = 0.10$ [63]. $T_c$ values are defined by criteria described in the text.

| Compound (Sample ID; pressure) | $T_\theta$ (K) | $T_c$ (K) | $\lambda_{e-ph,BCS}$ | $\lambda_{e-ph,aMcM}$ | $\lambda_{e-ph}$ (first-principles calculations) [63] |
|---|---|---|---|---|---|
| $\text{LaH}_x$ (panel **a**, Fig. 6) (Sample 10; $P = 178$ GPa) | $721 \pm 2$ | 92 | $0.586 \pm 0.001$ | $1.82 \pm 0.01$ | 2.06-2.76 |
| $\text{LaH}_x$ (panel **b**, Fig. 6) (Sample 10; $P = 178$ GPa) | $941 \pm 9$ | 107 | $0.560 \pm 0.002$ | $1.62 \pm 0.01$ | 2.06-2.76 |
| $\text{LaD}_{11}$ (panel **c**, Fig. 6) (Sample 8; $P = 142$ GPa) | $1199 \pm 14$ | 125 | $0.51 \pm 0.04$ | $1.49 \pm 0.02$ | 3.14 |

### 4.8.5. Other $\text{LaH}_x$ and $\text{LaD}_y$ samples

Drozdov *et al* [62] in their Extended Data Fig. 5 reported $R(T)/R_{\text{norm}}(T)$ curve for laser annealed $\text{LaD}_y$ (Sample 13) specimen compressed at $P = 152$ GPa, which has broad superconducting transition. Transition temperature can be estimated to be about $T_c = 125$ K if the inflection point criterion (Extended Data Fig. 5 [62]) will be applied. However, the fit of $R(T)/R_{\text{norm}}(T)$ to BG equation is not converged, and we were not able to report $T_\theta$ and $\lambda_{\text{e-ph}}$ values for this sample herein.

For similar problems, $T_\theta$ and $\lambda_{\text{e-ph}}$ cannot be deduced for two isotopic counterpart samples with highest reported transition temperatures, i.e. $\text{LaH}_{10}$ (Sample 1, $T_c \sim 250$ K) and $\text{LaD}_{10}$ (Sample 17, $T_c \sim 180$ K) [62].

### 5. Discussion

As it is mentioned above, one of fundamental problem in NRT superconductivity is, surprisingly enough, the definition of the transition temperature, $T_c$. Traditionally, the transition temperature in highly-compressed superconductors was defined at the onset of the $R(T)$ curve. This criterion is widely accepted and, for instance it was applied for highly compressed boron [26] and silane [28], and for many highly-compressed superconductors for which $R(T) = 0\ \Omega$ point has been never reached. However, if this criterion is accepted, then it should be applied for highly-compressed YH$_3$ [39], for which the ratio of $R(T)/R(T=270\ K) = 0.95$ gives $T_c = 170$ K (Fig. 1).

Thus, there is a need to acknowledge a fact that the superconducting state is established at the point of $R(T) = 0\ \Omega$, when the amplitude and the phase of the condensate are established within the whole sample. This means that at this point as the London penetration depth, $\lambda(T)$, as the coherence length, $\xi(T)$, have got finite values. Also, and this is crucially important, that the measurement of $R(T) = 0\ \Omega$ point is the simplest experimental measurement that can be done for sample inside of the diamond anvil cell.

The definition of the superconducting transition by the $R(T) = 0\ \Omega$ criterion is also important for the evaluation of the first principles calculations, because currently, due to many of highly-compressed superconductors have large width of the resistive transition (which can be defined by two temperatures at which $R(T)/R_{norm}(T) = 0.90$ and $R(T)/R_{norm}(T) = 0.10$ are reached), there is a problem to compare results of computed values with experiment.

However, based on a fact that Drozdov *et al* [5], Somayazulu *et al* [63], Troyan *et al* [31], and Kong *et al* [40] showed that $R(T) = 0\ \Omega$ state is possible to measure in highly-compressed hydrides at $T_c \geq 190\ K$ inside of the diamond anvil cell, then the definition of the superconducting transition by the $R(T) = 0\ \Omega$ criterion has a ground to be implemented for the whole field, including as experimental [5,26-28,31,32,40,45,47,51,52,56,58,62,68], as first principles calculations studies [6,7,9,29-32,37,50,53,58,60,63,70-73].
Returning now to the analysis performed in this paper, we should mention that first-principles calculations is a very accurate modern research tool, and large disagreement of computed ratio of logarithmic phonon frequencies (Eq. 31) and experimentally observed ratio of critical temperatures (Eq. 32) with deduced ratio of Debye temperatures (Eq. 33) we reported herein means that electron-phonon mechanism is irrelevant to the observed NRT superconductivity in $H_3S$ and $D_3S$, however it takes some effect of the second order on the magnitude of observed $T_c$. It is quite likely that electron-phonon mechanism is the origin for slightly higher transition temperature in $H_3S$ in comparison with $D_3S$, however, primary mechanism which is the background that both $H_3S$ and $D_3S$ are NRT superconductors is remains to be unknown.

It should be stressed, that the majority of applications of the BCS and the Eliashberg's theories to predict NRT superconductivity in hydrogen-rich highly-compressed compounds have been failed. For instance, we can mention the case of highly-compressed silane, $SiH_4$, for which $T_c = 98 - 107\ K$ was predicted by Li *et al* [74] and $T_c = 166\ K$ was predicted by Aschroft's group [75], while the experiment showed $T_c \leq 13\ K$ so far [28].

More thorough theoretical analysis of recent experimental milestone discoveries of NRT superconductivity in some hydride compounds [5,31,32,40,62,68] is required, because each successfully and each unsuccessfully predicted NRT superconductor cases should be treated with equal weight within total database, without fixing the database size by pre-defined stopping rule for unsuccessful cases. This approach will advance the progress in the field because it eliminates so-called survivorship bias [76,77].

There is also one fundamental problem on widely used Allen-Dynes approach [14,15] which has been never discussed, and which is the concept of logarithmic phonon frequency, $\omega_{ln}$ (Eq. 7). Truly, if one can consider the integrand part of the $\omega_{ln}$ definition (which is in square brackets):

$$\omega_{ln} = exp\left(\frac{\int_0^\infty \left[\frac{ln(\omega)}{\omega}\right] \cdot F(\omega) \cdot d\omega}{\int_0^\infty \frac{1}{\omega} F(\omega) \cdot d\omega}\right) = exp\left(\frac{\int_0^\infty \left[ln\left((\omega)^{\frac{1}{\omega}}\right)\right] \cdot F(\omega) \cdot d\omega}{\int_0^\infty \frac{1}{\omega} F(\omega) \cdot d\omega}\right) \qquad (37)$$

then the logarithm part of it:

$$ln(\omega) \qquad (38)$$

or even more complicated one:

$$ln\left((\omega)^{\frac{1}{\omega}}\right) \tag{39}$$

cannot be accepted to have physical meaning.

Any functions, i.e., *exp*(*x*), *cos*(*x*), modified Bessel function of $K_n(x)$, etc., and the logarithm as well, can be taken only from unit-less variable. However, in Allen-Dynes approach [14,15] the logarithm is taken from variable which has the unit of Hz (or the variable in units of Hz which is in power which has unit of seconds). There is simply not way to get mathematical and physical meaning for a value which raised in a power which has unit(s) of seconds, kilograms, meters, amperes, candela, etc.

All physical laws where oscillations are primary variable (for instance, Rayleigh's scattering law, Planck's law, Fourier transformation, etc.) the utilization of the frequency is unavoidably involved multiplicative pre-factors which eliminate the units of hertz.

Thus, a discussion why the concept of $\omega_{ln}$ (Eqs. 7,37) is so widely used as a valid approximation of Eliashberg's theory is required. Perhaps, the concept of $\omega_{ln}$ (Eqs. 7,37) is main reason why in so many cases predicted high $T_c$ values were never confirmed in experiment.

Alternative pairing mechanisms in superconductors are in discussion for several decades [48,49,61,64-67,78-86]. However, detail discussion of these more advanced approaches to understand NRT superconductivity as part of full picture of high-$T_c$ superconductivity is beyond the scope of this paper.

## 6. Conclusion

In this paper we deduce the Debye temperature, $T_\theta$, in highly-compressed black phosphorous, boron, GeAs, SiH$_4$, H$_x$S, D$_y$S, LaH$_x$ and LaD$_y$ by the fit of temperature-dependent resistivity data, $\rho(T)$, to Bloch-Grüneisen equation. We find that isotopic counterpart compounds (designated by subscripts 1 and 2) should obey the relation:

$$\frac{T_{c,1}}{T_{c,2}} = \frac{\omega_{ln,1}}{\omega_{ln,2}} = \frac{T_{\theta,1}}{T_{\theta,2}} \qquad (40)$$

which can be considered as a new research tool to validate the electron-phonon mechanism of superconductivity in a variety of the materials. Application of Eq. 40 to $H_3S$-$D_3S$ system leads us to a conclusion that NRT superconductivity in these compounds is originated from more than one mechanism.

**Acknowledgement**

Author thanks financial support provided by the state assignment of Minobrnauki of Russia (theme "Pressure" No. AAAA-A18-118020190104-3) and by Act 211 Government of the Russian Federation, contract No. 02.A03.21.0006.